%% file: article.tex
\documentclass[final,a4paper,12pt]{elsarticle}%\usepackage[utf8]{inputenc}
\usepackage[english]{babel}
\usepackage{amsmath}
\usepackage{amsfonts}
\usepackage{amssymb}
\usepackage{graphicx}
\usepackage[left=3cm,right=3cm,top=2cm,bottom=2cm]{geometry}
\usepackage{lineno}
\usepackage{caption} 
\usepackage{titlesec}
\usepackage{xcolor}
\usepackage{multicol}
\usepackage{adjustbox}
\usepackage{hyperref}
\usepackage[absolute,overlay]{textpos}

\makeatletter
  \long\def\pprintMaketitle{\clearpage
  \iflongmktitle\if@twocolumn\let\columnwidth=\textwidth\fi\fi
  \resetTitleCounters
  \def\baselinestretch{1}%
  \printFirstPageNotes
  \begin{center}%
 \thispagestyle{pprintTitle}%
  \def\baselinestretch{1}%
    \Large\@title\par\vskip18pt
    \normalsize\elsauthors\par\vskip10pt
    \footnotesize\itshape\elsaddress\par\vskip36pt
    \end{center}%
  \gdef\thefootnote{\arabic{footnote}}%
  }
  
\def\ps@pprintTitle{%
  \let\@oddhead\@empty
  \let\@evenhead\@empty
  \let\@oddfoot\@empty
  \let\@evenfoot\@oddfoot
  }

\makeatother
\newcommand{\myModifColor}{black}
\newcommand{\PRCC}{black}
\newcommand{\REV}{black}
\newcommand{\REVISION}{black}
\newcommand{\MAU}{black}
\newcommand{\LASTVR}{black}

\date{}

 \usepackage{float} 
 \usepackage{wrapfig}

\begin{document}
%\addauthor
\input{other/author_list_Nuclearites_06072022}

\input{other/abstract}

\title{\boldmath Limits on the nuclearite flux using the ANTARES neutrino telescope}
\maketitle

%\input{abstract}
  %  \linenumbers
   % \newpage
      %  \input{abstract} 

    \newpage
        \input{sections/introduction}
        \input{sections/theAntaresNeuTele}

        \input{sections/nuclearites}
        \input{sections/nuclearitesSimulation}

        \input{sections/eventSelection}

        \input{sections/resultsAndDiscusion}
        \input{sections/conclusion}
        \input{other/Ackn}

    \newpage

\input{sections/bibliography}
\end{document}

%% file: other/author_list_Nuclearites_06072022.tex
%\documentclass{elsarticle}
%\journal{}
%\begin{document}
%\begin{frontmatter}
%\title{ANTARES collaboration}
\author[IPHC,UHA]{A.~Albert}
\author[IFIC]{S.~Alves}
\author[UPC]{M.~Andr\'e}
\author[UPV]{M.~Ardid}
\author[UPV]{S.~Ardid}
\author[CPPM]{J.-J.~Aubert}
\author[APC]{J.~Aublin}
\author[APC]{B.~Baret}
\author[LAM]{S.~Basa}
\author[CNESTEN]{B.~Belhorma}
% ...

\address{(ANTARES Collaboration)}

\address[IPHC]{\scriptsize{Universit\'e de Strasbourg, CNRS,  IPHC UMR 7178, F-67000 Strasbourg, France}}
\address[UHA]{\scriptsize Universit\'e de Haute Alsace, F-68100 Mulhouse, France}
\address[IFIC]{\scriptsize{IFIC - Instituto de F\'isica Corpuscular (CSIC - Universitat de Val\`encia) c/ Catedr\'atico Jos\'e Beltr\'an, 2 E-46980 Paterna, Valencia, Spain}}
\address[UPC]{\scriptsize{Technical University of Catalonia, Laboratory of Applied Bioacoustics, Rambla Exposici\'o, 08800 Vilanova i la Geltr\'u, Barcelona, Spain}}
\address[UPV]{\scriptsize{Institut d'Investigaci\'o per a la Gesti\'o Integrada de les Zones Costaneres (IGIC) - Universitat Polit\`ecnica de Val\`encia. C/  Paranimf 1, 46730 Gandia, Spain}}
\address[CPPM]{\scriptsize{Aix Marseille Univ, CNRS/IN2P3, CPPM, Marseille, France}}
\address[APC]{\scriptsize{Universit\'e de Paris, CNRS, Astroparticule et Cosmologie, F-75013 Paris, France}}
\address[LAM]{\scriptsize{Aix Marseille Univ, CNRS, CNES, LAM, Marseille, France }}
\address[CNESTEN]{\scriptsize{National Center for Energy Sciences and Nuclear Techniques, B.P.1382, R. P.10001 Rabat, Morocco}}
% ...
%\end{frontmatter}
%\end{document}

%% file: other/abstract.tex
\if false
In this paper, we report the results obtained for down-going flux of nuclearites using nine years (i.e. from 2009 to 2017) of data taking with ANTARES neutrinos telescope. We used the model described by A. De $R\hat{u}jula$ and S. L. Glashow to simulate the nuclearites with a galactic velocity of $ \beta = 10^{-3} $ at the top of the atmosphere. Two discrimination variables were used in order to extract the signal from the background events due mainly to atmospheric muons and bioluminescence. Particular cuts were used to clean the data and to compute the nuclearites flux upper limit in ANTARES for the considered period of data taking. The results are showing a good improvement with regards to previous reported results.
\vfill
\fi

\begin{abstract}
In this work, a search for nuclearites of strange quark matter by using nine years of ANTARES data taken in the period 2009-2017 is presented. The passage through matter of these particles is simulated %according to the model of de R\'{u}jula and Glashow 
taking into account a detailed description of the detector response to nuclearites and of the data acquisition conditions. A down-going flux of cosmic nuclearites with Galactic velocities ($\beta = 10^{-3}$) was considered for this study. The mass threshold for detecting these particles at the detector level is \mbox{ $4 \times 10^{13}$ GeV/c$^{2}$}. Upper limits on the nuclearite flux for masses up to $10^{17}$ GeV/c$^{2}$ at the level of $\sim 5 \times 10^{-17}$ cm$^{-2}$ s$^{-1}$ sr$^{-1}$ are obtained. These are the first upper limits on nuclearites established with a neutrino telescope and the most stringent ever set \textcolor{\PRCC}{for Galactic velocities}.
\end{abstract}

%The results obtained are improving the previous published results. Therefore, a new limit on the flux of nuclearites of strange matter is set for the explored region.

\if false

Atmospheric muons events constitute the main background for this analysis, other form of incoherent noise presenting in the experimental data related to the detection medium rises from bioluminescence and $^{40}K$ decay.

 by more than one order of magnitude

Intermediate mass objects of strange quark matter (SQM), nuclearites, may exists in the cosmic rays reaching the Earth, and they would leave a distinct signal in the cosmic rays detectors. ANTARES, an array of photomultipiers (PMTs), is a Cherenkov based telescope located at about 2.5 Km under water in the Mediteraninien sea, it would be able to detect nuclearites according to the model of A. De $R\hat{u}jula$ and S. L. Glashow. In this work, we consider a down-going flux of nuclearites penetrating to the Earth with galactic velcities ($\beta = 10^{-3}$) to search for the signal of these particles in ANTARES data. Also we present the analysis performed and the results obtained for a period of nine years of data taking (from 2009 to 2017).
\fi

%% file: sections/Introduction.tex
\section{Introduction}
\par

\if false 
It has been suggested for about 36 years, that strange quark matter (SQM) could be the true ground state for hadronic matter [\ref{bib:EWitten}]. SQM is a new form of matter proposed by Witten in 1984 in which, up, down and strange quarks could be confined in a structure commonly called bag with comparable numbers. The hadronic bag could be as small as ordinary nuclei or as large as a star A $\sim 2 \times 10^{57}$. Macroscopic quark matter lumps surviving from the cosmological quark-hadron phase transition are often referred to as quark nuggets, and if they manage to reach the Earth they are called nuclearites [\ref{bib:EWitten}, \ref{bib:FWeber}, \ref{bib:Farhi&Jaffe}].

Nuclearites are hypothetical heavy particles derived from the SQM Theory. They could be present in the cosmic radiation reaching the Earth originating from relics of the early Universe as nuggets or strange star collisions. They could be detected by using the light generated by their atomic collision as they cross transparent mediums [\ref{bib:Glashow&DRegula}].

In this work, we study cosmic nuclearites falling on the Earth with galactic velocities \mbox{($\beta = 10^{-3}$)}, we conseder a mass of $4 \times 10^{13} \ GeV/c^2$ as a threshold mass for nuclearites detection at ANATRES level. A temporal resolution of $2ns$ were used in the simulation software. Four nuclearites masses were produced: $4 \times 10^{13} , 10^{14}, 10^{15}$ and $10^{16} \ GeV/c^2$ by using a dedicated MonteCarlo simulation software.

Nuclearites are hypothetical heavy particles derived from the strange quark matter (SQM) Theory \cite{EWitten}. They could be present in the cosmic radiation reaching the Earth originating from relics of the early Universe as nuggets or strange star collisions. They could be detected by using the light generated by their atomic collision as they cross a transparent mediums \cite{DeRujula&Glashow,Popa}.

In this work, we study cosmic nuclearites falling on the Earth with galactic velocities $\beta = 10^{-3}$, assuming a mass of $4 \times 10^{13} $  GeV/c$^2$ as a threshold mass for nuclearites detection at the ANATRES level. By using a dedicated Monte Carlo simulation, each particle is generated with random zenith and azimuth angles and its velocity is evaluated, then propagated throw the detector where the signals are produced. A first results on sensitivities of ANTARES to these particles using data collected in the period 2009-2017 is presented.
\fi

%First introduced by E. Witten, strange quark matter (SQM) \cite{Bodmer,Terazawa,EWitten} is a new form of matter that could be more stable than ordinary matter. 
The idea of strange matter goes back to 1971 when A. Bodmer discussed collapsed nuclei \cite{Bodmer}. In 1984 E. Witten introduced a new form of matter named strange quark matter (SQM) that could be more stable than ordinary matter \cite{EWitten,Terazawa}. SQM is made up of roughly equal numbers of 
\textcolor{\MAU}{up, down and strange quarks with color neutrality. SQM objects are confined in a hadronic structure commonly called “bag”\cite{EFarhiRLJaffe}. A bag of SQM will have a net positive charge on the surface with an almost equal electrostatic potential of + 50 MeV. This potential is adequate to repel positively charged nuclei at ordinary temperatures or at reasonably low velocities, and to attract electrons to neutralize the electric charge of SQM bags \cite{EWitten}}.
%up, down and strange quarks with electric and color charge neutrality. SQM objects are confined in a hadronic structure commonly called ``bag"
Bags can be as small as ordinary nuclei (baryon number $A\lesssim 260$) or as large as a neutron star (baryon number A $\simeq 2 \times 10^{57}$) \cite{EWitten,Bauswein}. SQM has been studied as a part of the MIT\footnote{``MIT bag model” named after the institution (Massachusetts Institute of Technology
) of the inventors of the model.} bag model for different strange quark masses and bag constants \cite{EFarhiRLJaffe,Chodos,DeGrand}. It could exist if it is a preferred thermodynamic phase compared to the normal state of nuclear matter consisting of protons and neutrons \cite{Greiner}. SQM could be formed either during the quark-hadron phase transition in the early Universe, or during the conversion of neutron stars into strange stars where strange quarks would be produced through weak processes \cite{Bauswein,Bjorken}.

SQM \textcolor{\myModifColor}{produced} in the early Universe was originally  suggested as a candidate for dark matter \cite{EWitten,DeGrand}. A. \textcolor{\myModifColor}{D}e R\'{u}jula and S. L. Glashow have discussed methods to detect the Galactic flux of SQM objects falling on Earth and used the available data to constrain the cosmic flux of SQM \cite{DeRujulaGlashow}. 
%Macroscopic quark matter lumps surviving from the cosmological quark-hadron phase transition are often referred to as quark nuggets, and if they manage to reach the Earth, they are called nuclearites. Cosmic nuclearites would be expected to have galactic velocities and could be identified by tracks seen in cosmic ray detectors based on the Earth 
Usually \textcolor{\myModifColor}{in the literature}, small ($A < 10^7$) SQM systems in the GeV-TeV range are called \textit{strangelets}%, they are searched for in collider experiments (eg. MoEDAL at LHC)
. Strangelets are outside \textcolor{\PRCC}{the range of neutrino telescopes}: a strangelet is searched for as an event with anomalous charge-to-mass ratio in the cosmic radiation using balloon or space-borne spectrometers. However, neutrino telescopes might detect nuclearites that penetrate at their depth. The term \textit{nuclearites} is used to design higher mass ($A > 10^7$) objects. Nuclearites are electrically neutral atom-like systems, as they would be expected to possess an electron cloud around the core. For $A > 10^{15}$, electrons would be largely contained within the bag of nuclear matter. 

Nuclearites are generally assumed to be \textcolor{\REV}{gravitationally} bound to astrophysical objects (the Galaxy, the Galaxy cluster...) and to have a speed determined by the virial theorem. In the case of nuclearites bound to the Milky Way, their velocity relative to the Solar System reference frame is $\beta \sim 10^{-3}$. Nuclearites can arrive at the Earth ground and penetrate large amounts of material, reaching underground detectors. Usually, they are searched for by cosmic ray detectors \cite{Madsen, Maurizio}. These particles are protected against direct interactions with the atoms constituting the traversed medium by an electronic cloud surrounding their core. However, they would lose a large amount of energy by elastic and quasi-elastic collisions with the atoms encountered in their path. Hence, they leave a distinct signal during the passage in cosmic ray detectors. In the case of a transparent medium such as water, nuclearites signal could be detected by using the light emission from their overheated path as a black-body radiation from an expanding cylindrical thermal shock wave \cite{DeRujulaGlashow}.

This work considers downgoing cosmic nuclearites arriving on the Earth with velocities \mbox{$\beta = 10^{-3}$}. In order to determine the threshold detection mass of these particles at the detector level, nuclearites with different masses are assumed \textcolor{\myModifColor}{to arrive vertically into} the detector. The arrival speed of nuclearites at the detector depth is computed after considering their energy loss in atmosphere and sea water. Preliminary results obtained with a small ANTARES data set equivalent to 159 days of data \textcolor{\myModifColor}{were} presented in \cite{Pavalas}.
%by using the formulas of energy loss in atmosphere and sea water, we computed the arrival speed at the detector depth, and so, the number of visible photons expected per unit of length using formulas given in section \ref{sec:nuclearites}.

 %Particles with a mass of $9 \times 10^{12}$ GeV/c$^2$ are producing about 1000 photons/cm at the ANTARES depth.
%This is only marginally above the light yield from atmospheric muons. Therefore nuclearites of such a low mass cannot be detected in ANTARES.
 %However, the path of the particles is not necessarily vertical and the detector response is affected by the environmental noise. We have simulated 100 nuclearites events with a mass of $3 \times 10^{13}$ and the results showed a low rate of events that could not be used in an analysis to set a limit on the flux. 
 %Therefore, we considered a mass of $4 \times 10^{13}$ GeV/c$^2$ as a threshold mass for nuclearites detection at ANTARES depth as it provides a sufficient light yield for an unambiguous detection.. 

%For this analysis, atmospheric muons constitute the main source of background events, other form of noise rises from bioluminescence and $^{40}K$ decay in the real data, this noise constitute a real challenge toward the signal isolation since it exhibit the same behaviour as nuclearites. A Monte Carlo simulation for both nuclearites and atmospheric muons signals in ANTARES has been made, this simulation takes into account the real data acquisition conditions.

In this paper, a brief description of the ANTARES neutrino telescope is given in section \ref{sec:Antares}. Section \ref{sec:nuclearites} introduces the main nuclearite properties, in particular their effective cross-sectional area and the light yield in a transparent medium. The simulation of the passage of nuclearites in the ANTARES detector and the \textcolor{\myModifColor}{associated} event selection are discussed in sections \ref{sec:simulation} and \ref{sec:eventSelection}, respectively. The results obtained after the analysis of 9 years of data are presented and discussed in section \ref{sec:resultsAndDiscussion}. %Section \ref{sec:conclusion} contains the conclusions.
%then, in section \ref{sec:nuclearites} the nuclearites as well as the formulas that governs the interactions of these particles are introduced. The simulation and the events selection is discussed in section \ref{sec:simulation} and in section \ref{sec:eventSelection}. The results obtained are showed and discussed in section \ref{sec:resultsAndDiscussion} and we give a conclusion on this work in section \ref{sec:conclusion}.

%% file: sections/theAntaresNeuTele.tex
\section{The ANTARES neutrino telescope}
\label{sec:Antares}

ANTARES (Astronomy with a Neutrino Telescope and Abyss environmental RESearch) was a neutrino telescope based on the detection of Cherenkov radiation whose construction was completed in May 2008. The data acquisition was definitively stopped on February, 12th 2022 and the detector decommissioning \textcolor{\myModifColor}{ended in June} 2022. It was made up of 12 vertical detection lines of 450 m length each, horizontally spaced by about \textcolor{\myModifColor}{$60-75$} m. A floor (or a storey) was formed by three optical modules (OMs) housing a 10" photomultiplier tube (PMT). A detection line was an ensemble of 25 floors spaced by 14.5 m \textcolor{\myModifColor}{being} the first one located 100 m from the seafloor. The detector was deployed with the base of the lines at a depth of \textcolor{\PRCC}{$\sim$}2475 m in the Mediterranean Sea, 42 km offshore from Toulon in France. The lines were anchored to the seabed with a dead weight and were held vertical by a buoy at the top. The so-called junction box supplied the lines with electrical power and bundles the data streams from the OMs as well as the distribution of control signals to the various components. % \ref{Fig:Antares} shows a schematic view of the ANTARES detector and 
\textcolor{\PRCC}{When photons of any origin impinge on the PMT photocathode, a signal can be produced at the anode and converted into a digital format by the front-end electronics boards \cite{ARS}, recording time and charge, and storing information in what is called a \textit{hit}}. A detailed description of the apparatus \textcolor{\PRCC}{can} be found in \cite{Aguilar}.

Downward going atmospheric muons and atmospheric neutrinos interacting inside or in the vicinity of the detector \textcolor{\PRCC}{produce} the bulk of physics signal. In fact, relativistic charged particles passing in or close to the telescope induce \textcolor{\myModifColor}{the production of} Cherenkov photons in the medium\textcolor{\myModifColor}{, which} can be detected by the PMTs. Nuclearites, as heavy penetrating particles, would generate visible photons from a different mechanism \textcolor{\myModifColor}{(described in the next section)} while they penetrate into water, producing a characteristic and recognizable signal in the ANTARES detector.

%% file: sections/nuclearites.tex
\section{Interaction of nuclearites with matter}
\label{sec:nuclearites}
%\paragraph{}
Nuclearites are assumed to be heavy strange quark nuggets. %The global neutrality is ensured by an electron cloud surrounding the nuclearite core forming a sort of an atom. Nuclearites with galactic velocities are protected by their surrounding electrons and Coulomb repulsion against direct interactions with the atoms constituting the traversed medium. Thus, the principal energy-loss mechanism for a nuclearite passing through matter is atomic collision.
For a massive nuclearite the energy loss is mainly due to the elastic and quasielastic collisions with atoms constituting the traversed medium. The energy loss per unit of length is \cite{DeRujulaGlashow}

\begin{equation}
-\dfrac{dE}{dx} = \rho \sigma v^2 ,
\end{equation}
\noindent 
where $\rho$ is the density of the traversed medium, $v$ the non-relativistic velocity of the nuclearite and $\sigma$ the effective \textcolor{\myModifColor}{cross section}  of the nuclearite defined as \cite{DeRujulaGlashow}

\begin{equation}
\label{equ:crossSection}
\sigma \textcolor{\myModifColor}{[\text{cm}^2]} =
\begin{cases}
\pi \times 10^{-16} & \text{if} \quad M_{N} < \textcolor{\myModifColor}{1.5 \times 10^{-9} \ \text{g} } \\
\pi \times \bigg(\dfrac{3M_{N}}{4 \pi \rho_{N}} \bigg)^{2/3} & \text{if} \quad M_{N} \geq \textcolor{\myModifColor}{ 1.5 \times 10^{-9} \ \text{g} }
\end{cases},
\end{equation}
where $M_{N}$ is the nuclearite mass and $\rho_{N} = 3.6 \times 10^{14}$ g/cm$^3$ \textcolor{\PRCC}{is} the density of SQM objects \textcolor{\REV}{\cite{Chin}}. In Eq. \ref{equ:crossSection} the mass limit corresponds to a nuclearite with a radius of about $10^{-10}$ m. As the chemical potential of the \textcolor{\PRCC}{\textit{s}} quark in SQM is slightly larger than for \textcolor{\PRCC}{\textit{u}} and \textcolor{\PRCC}{\textit{d}} quarks, finite SQM objects are always positively charged \textcolor{\REV}{and this charge is neutralized by an electron cloud surrounding their core}, thus the effective \textcolor{\PRCC}{cross section} for nuclearites with $M_{N} < 1.5 \times 10^{-9}$ g is controlled by their electronic cloud.

Nuclearites would travel with the typical velocity of
gravitationally trapped objects in our Galaxy. Therefore, nuclearites are assumed entering the Earth\textcolor{\PRCC}{'s} atmosphere with a velocity of $\beta_{0}=10^{-3}$. Before they reach the ANTARES detector they propagate through a large amount of material and interact with its constituents. %At the ANTARES detector, nuclearites should have a velocity of $\beta_{L} < \beta_{0}$ 
%and then continue propagating toward the ANTARES sensitive volume throw water where the density was taken $\rho_{w} = 1 \ g \ cm^{-3}$
For nuclearites of mass $M_N$ penetrating a distance $L$ in a medium of density profile $\rho(x)$, \textcolor{\myModifColor}{their} velocity changes as \cite{DeRujulaGlashow}

\begin{equation}
\beta_{L}=\beta_{0} \times \exp \left( {-\dfrac{\sigma}{M_N} \int_0^L \rho(x) dx } \right).
\label{equ:betaVaration}
\end{equation}

\textcolor{\myModifColor}{To simulate the propagation of} nuclearites along the Earth\textcolor{\PRCC}{'s} atmosphere, the following parametrization  of its density is used \cite{Shibata}\textcolor{\PRCC}{:}

\begin{align}
\label{equ:atmospher}
\rho(x) = a \times \exp \left(-\dfrac{x}{b} \right) = a \times \exp \left(-\dfrac{H-L(x)}{b} \right),
\end{align}
where $a=1.2 \times 10^{-3}$ g/cm$^{3}$, $b = 8.57 \times 10^5$ cm, $H = 50$ km is the height of the atmosphere and $L(x)$ is the penetrating length in the atmosphere. \textcolor{\myModifColor}{For the propagation in water, a constant density of $\rho(x)$ = 1 g/cm$^{3}$ is assumed.}
%In the case of sea water, the density was taken as $\rho_{w} = 1 \ g \ cm^{-3}$.

%The nuclearites signal could be detected through their visible light emission in the form of black body radiation from an expanding cylindrical thermal shock wave [2], the fraction of energy transformed in light called luminous efficiency $\eta$ was estimated to be $\eta \simeq 3 \times 10^{-5}$ in the case of water; thus,

Any experimental search for nuclearites has an acceptance of the detector for an isotropic flux of nuclearites that depends on their mass. Only nuclearites with sufficiently large mass ($>2.5 \times 10^{22}$ GeV/c$^2$) can traverse the Earth at typical Galactic velocities \cite{Wan-Lei}. As the flux of nuclearites should decrease when their mass increases, in this study\textcolor{\myModifColor}{,} downward going nuclearites with masses lower than $10^{17}$ GeV/c$^2$ are considered. In fact, downward going nuclearites of larger masses are expected to \textcolor{\myModifColor}{cause} an overall saturation of the detector, producing events with a high number of hits that could not be processed. %almost indistinguishable from a detector malfunctioning. 

A summary of the detection techniques and experimental results can be found in \cite{MACRO, SLIM}. Transparent media (e.g., liquid scintillators or water) have been used in \textcolor{\myModifColor}{nuclearite} searches. Nuclearites\textcolor{\PRCC}{,} as non-relativistic objects\textcolor{\PRCC}{,} do not produce Cherenkov light. They \textcolor{\myModifColor}{are expected to} give rise to a thermal shock through collisions with the atoms of water. The temperature of the medium surrounding the nuclearite path rises \textcolor{\myModifColor}{up to the order of a few} \textcolor{\REV}{thousands of \textcolor{\MAU}{kelvin}}. Thus, a hot plasma is formed that moves outward as a shock wave, emitting blackbody radiation and producing many photons in the visible band. A detailed description of the luminous efficiency $\eta$ \textcolor{\myModifColor}{defined as the fraction of dissipated energy appearing as light} is given in \cite{DeRujulaGlashow}. The authors estimate that in pure water, a fraction of about $3\times 10^{-5}$ 
\textcolor{\MAU}{of the total energy loss is provided in form of photons in the energy range [2.25-3.75] eV. This region is smaller than the wavelengths range in which ANTARES PMTs are sensitive (300nm-600nm). The number of visible photons radiated per unit of path length is conservatively estimated as
}

\begin{align}
\dfrac{dN_\gamma}{dx} = \dfrac{\eta}{\pi} \times \dfrac{dE/dx}{\text{eV}}.
\label{eq:nbVisPhotons}
\end{align}

\textcolor{\MAU}{
The possible theoretical uncertainty on the luminous efficiency $\eta$ exactly in the 300-600 nm band is within the systematic uncertainties arising from  water transparency and PMT detection efficiency variation as a function of time, as discussed in sect. \ref{sec:sysErr}
}
%of the total energy loss is provided in form of visible light. The number of visible photons radiated per unit of path length is estimated as

% \begin{align}
% \dfrac{dN_\gamma}{dx} = \dfrac{\eta}{\pi} \times \dfrac{dE/dx}{\text{eV}}.
% \label{eq:nbVisPhotons}
% \end{align}

% The intensity $I$ for an isotropic source of photons with intensity $I_{0}$ detected at a distance $r$ from the source is

% \begin{align}
% I = I_{0} \times \dfrac{1}{4 \pi r^{2}} \times \exp \bigg(\dfrac{-r}{\lambda_{att}} \bigg).
% \end{align}

At the depth of the ANTARES detector and under the previous conditions, the number $I$ of photons reaching one photomultiplier tube is given by the integration of the number of visible photons radiated per unit of length (Eq. \ref{eq:nbVisPhotons}) over the nuclearite path length into the detector\textcolor{\myModifColor}{. The expression is given by}
%the photomultiplier output signal in the case of a photon with a wavelength $\lambda$ and an incidence angle $\theta$ is given by

%\dfrac{dS_{\gamma}}{dx} (\lambda, \theta) 

\begin{align}
I = \int \dfrac{\Omega}{4 \pi} \dfrac{dN_\gamma}{dx} \times \exp \bigg(\dfrac{-r}{\lambda_{\text{att}}}\bigg) dx, 
\end{align}
%I \propto  N_{\gamma}(\lambda)

%I= N_\gamma \times QE(\lambda) \times A_{Acc}(\theta) \times \dfrac{A_{eff}}{4 \pi r^{2}} \times \exp \bigg(\dfrac{-r}{\lambda_{att}(\lambda)}\bigg).

%$N_{\gamma}$ is the number of visible photons isotropically emitted along the nuclearite path, $r$ is the distance from the nuclearite position to the PMT in consideration,

\noindent
\textcolor{\myModifColor}{where $\Omega$ is the solid angle in which the optical module is seen from the emission point defined as}

\begin{align}
\Omega = \dfrac{ A_{\text{eff}} \times \cos(\theta) }{r^{2}},
\end{align}

\noindent 
\textcolor{\myModifColor}{
\textcolor{\PRCC}{where} $r$ is the distance from the nuclearite position to the PMT and  $\theta$ is the incidence angle of the photon on the PMT. The quantity $\lambda_{\text{att}}$ is the light attenuation length in sea water. $A_{\text{eff}}$ is the effective area of the PMT \cite{Aguilar}. \textcolor{\PRCC}{T}he number of photoelectrons in the PMT is computed including the quantum efficiency of the tubes.
}
% 4.11%  4.41% 4.39% 6.30%

%% file: sections/nuclearitesSimulation.tex
\section{Monte Carlo simulation}
%\subsection{Real data and simulation}
\label{sec:simulation}

\begin{wrapfigure}{hr}{0.45\textwidth}
\centering
\includegraphics[width=0.49\textwidth]{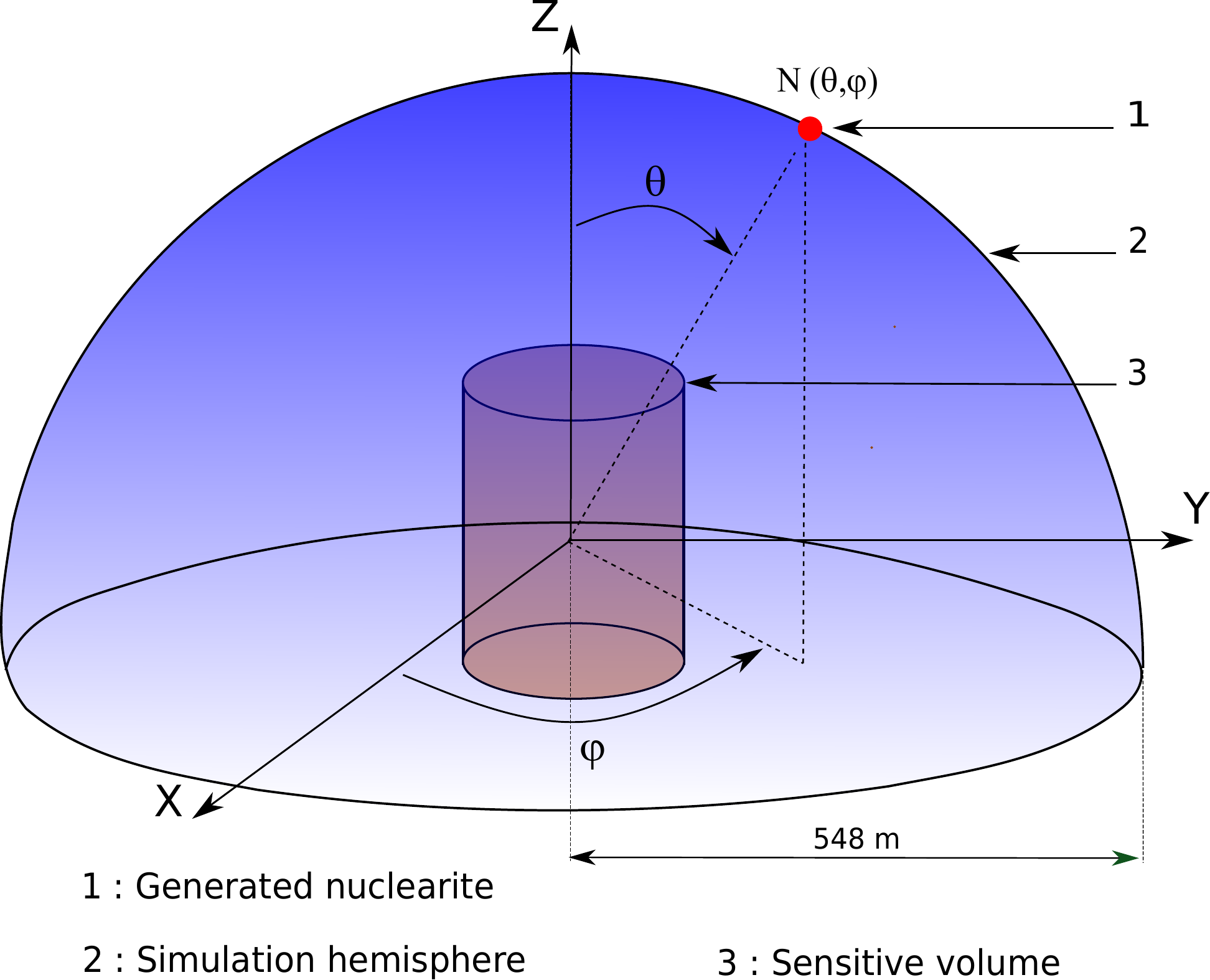}\label{two}
\caption{Simulated geometries for the event generation. \textcolor{\myModifColor}{Nuclearites are generated with random position over the surface of the hemisphere, isotropic randomly generated zenith and azimuth angles $(\theta,\varphi)$ that define the particle direction. \textcolor{\PRCC}{They are} propagated \textcolor{\PRCC}{and o}nly those entering in the detector sensitive volume are then considered in the simulation.}
}
\label{fig:simuHemi}
\end{wrapfigure}

To simulate \textcolor{\myModifColor}{downward} going nuclearites arriving at the ANTARES telescope, a hemisphere of 548 m radius has been used as a generation volume filled with water, \textcolor{\myModifColor}{with the value of the radius corresponding to} two light attenuation lengths in water from the closest storey. The considered medium surrounds symmetrically the instrumented volume of the detector, as shown in Fig. \ref{fig:simuHemi}. Each simulated event consists of a nuclearite generated with random position over the surface of the generation volume and with \textcolor{\myModifColor}{isotropic randomly generated} zenith and azimuth angles $(\theta,\varphi)$ that \textcolor{\myModifColor}{define} the particle direction. Since the nuclearites are assumed to arrive at the Earth with Galactic velocities,
%we should calculate the velocity of the particle at the generation point; this is done by inverse calculation of the path crossed in both water and atmosphere using the 
\noindent
 the velocity of the  particle at the generation point is determined  using the path crossed in \textcolor{\myModifColor}{sea} water and atmosphere \textcolor{\PRCC}{according to} Eq. \ref{equ:betaVaration}. At this stage, the particle is propagated along its direction with \textcolor{\myModifColor}{a time step} of 2 ns. \textcolor{\myModifColor}{The} position and the speed are evaluated at each iteration as well as the energy loss, the number of expected visible photons and the number of \textcolor{\myModifColor}{photoelectrons produced in each OM}, which are calculated using the formulas introduced in section \ref{sec:nuclearites}. The simulation ends when the particle exits the simulation hemisphere or \textcolor{\myModifColor}{when} its optical energy loss (integrated over the time step) becomes too small to produce a sufficient amount of visible photons (\textcolor{\myModifColor}{i.e. it drops below 3 eV}). \textcolor{\myModifColor}{At the end, the analysis tool simulate\textcolor{\PRCC}{s} the  digitization of the signal on PMTs, yielding hits as for real events}. 
 
 %the QE and the OM collection efficiency corrections are derived from the $^{40}$K decay measurements.

Atmospheric muons represent the major background source for \textcolor{\myModifColor}{this analysis}. They are produced in the decay of charged mesons generated by the interactions of primary cosmic rays with nuclei \textcolor{\myModifColor}{present} in the atmosphere. %Downward going atmospheric muons events constitute the main background for nuclearites. 
\textcolor{\myModifColor}{In the simulations,} they are generated in \textcolor{\myModifColor}{bundles and} propagated down to 5 km water equivalent with the MUPAGE event generator \cite{mupage}. MUPAGE is based on parametric formulas allowing to calculate the flux and angular distribution of underwater or ice muon bundles. The generator takes into account the muon multiplicity and the multi-parameter dependent energy spectrum.

%For this analysis, we used atmospheric muons generated with the MUPAGE code \cite{mupage}. Other background events come from bio-organisms and the decays of potassium 40 ($^{40}$K) with a rate of about 34 kHz.
The last step of the simulation chain aims at transforming the individual \textcolor{\myModifColor}{light pulses} into a data stream with the same format and environmental conditions as real data. In order to meet this objective, the environmental optical background due to bio-organisms and the decays of $^{40}$K \textcolor{\myModifColor}{are} added to the light produced by physics events (nuclearites or atmospheric muons). Also the behaviour of individual optical modules can be affected by local changes of environmental conditions. As a consequence, the time evolution of the data acquisition \textcolor{\myModifColor}{is} properly reproduced, as described in \cite{mc}.

%The detector response was simulated using the TriggerEfficiency program. This software reproduces, by using a true run, the real data acquisition conditions for each run, also it adds the contribution of hits from bioluminescence and $^{40}$K decay.
% as it was during the data record.v

%% file: sections/eventSelection.tex
\section{Events selection and analysis}
\label{sec:eventSelection}

In order to avoid biased results, a fraction of 10\% of real data has been used \textcolor{\myModifColor}{to define the event selection criteria} and to compute the sensitivity of the detector to nuclearites. %However, the analysis is very sensitive to the environmental optical background which presents the same behaviour as a nuclearite and this complicates the isolation of our signal. As consequence, 
%Strong selection criteria are applied for the runs samples in order to select a clean dataset where only runs with a baseline rate of the optical background is lower than 120kHz and the burst fraction is below 40\% are selected; from nine years of data, we have selected about 839 days of livetime of the detector. %Besides, we use the 
\textcolor{\PRCC}{In addition, as} the detection of nuclearite signals could be impacted by bioluminescence, a strict anti-bioluminescence cut is applied. A selection is made for each data taking period of several hours, which is called ``run". To be considered for this analysis, a ``run" should not have more than 20\% of the detector elements affected by bioluminescent bursts at any moment. After applying this filter, 839 days of livetime in the period 2009-2017 are selected.

 %to reduce the optical noise presenting in the real data, a preliminary quality cut requiring at least 300 L0 hits for each event has been applied. This cut, also allows to remove high energies atmospheric muons events characterized by relatively high amplitude and low number of floors

\subsection{Trigger and event selection}
\label{Sec:sec5_1}

All hits with a charge above a minimum threshold of 0.3 photoelectrons (p.e.) are denoted as L0 \textcolor{\myModifColor}{hits} and transmitted to the shore and processed with dedicated trigger algorithms to identify potentially interesting events that are stored on disk \cite{DAQ}. L1 hits are defined as either a coincidence of two L0 hits \textcolor{\myModifColor}{in} the same storey within a time window of 20 ns or a single hit with high amplitude exceeding a predefined high-threshold condition (set to 3 p.e. or 10 p.e. depending on the data acquisition conditions).

ANTARES \textcolor{\REV}{used} several trigger algorithms to filter its data. For this analysis, the standard muon triggers \textit{T3} and \textit{3D} are used to \textcolor{\myModifColor}{characterise} the \textcolor{\myModifColor}{nuclearite} signal. %in order to characterize the signal.  
The 3D trigger requires at least 5 causally connected L1 hits %is a directional trigger, it checks for hits causally connected along the same trajectory, it requires at least 5 L1 hits (see section \ref{sec:Antares})
 within 2.2 $\mu$s of each other and the T3 trigger is defined as the occurrence of at least two L1 hits in three consecutive storeys within a coincidence time window. This coincidence time window is set to 100 ns in the case that the two storeys are adjacent, and 200 ns in the case of next to adjacent storeys.

\begin{figure}[htbp]
\centering
\includegraphics[width=.8\textwidth, height=8cm]{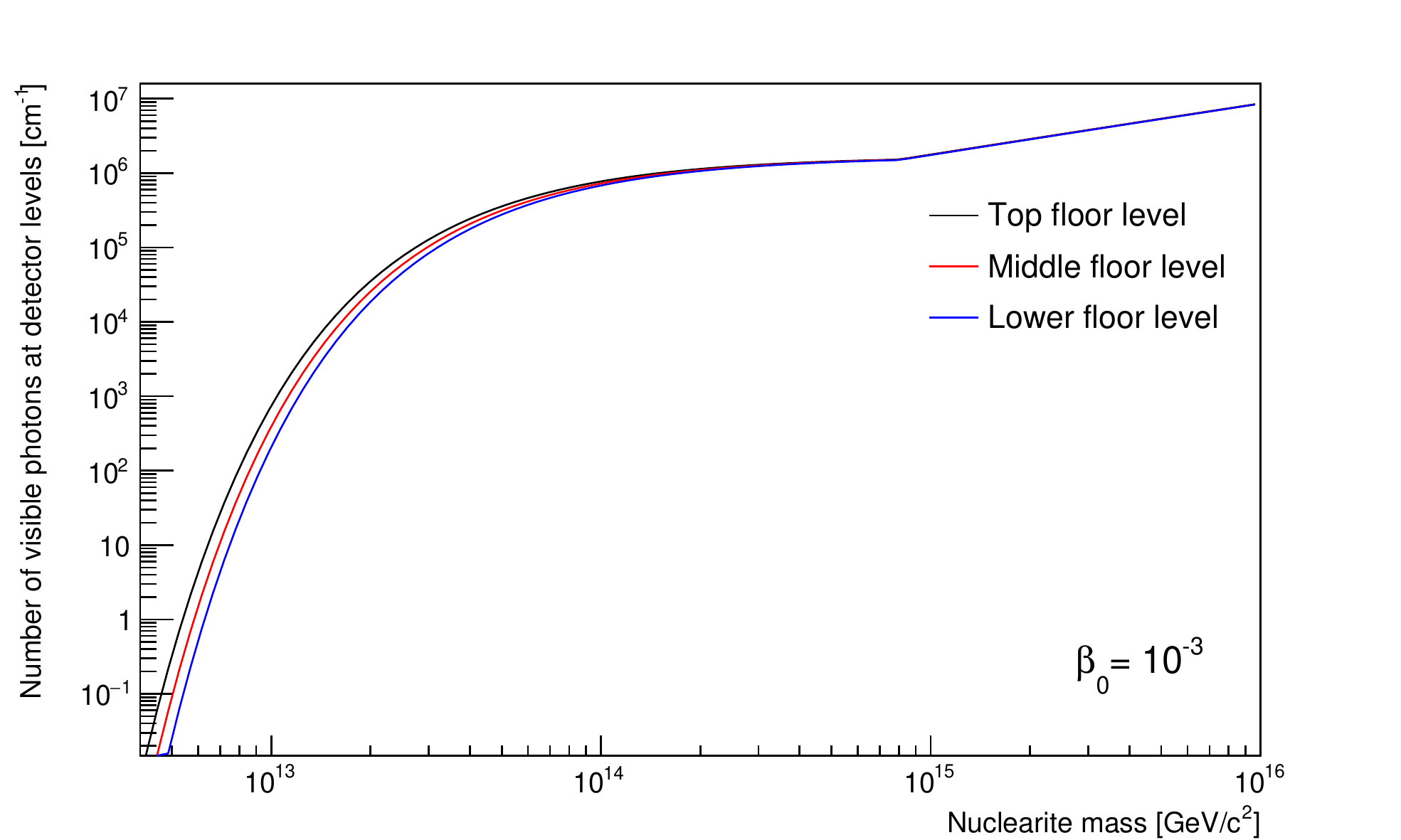}

\caption{Number of visible [300-600 nm] photons per centimeter generated by vertically incident nuclearites at different detector levels. For masses higher than $8.4 \times 10^{14}$ GeV/c$^2$ all electrons must be inside the quark bag. \textcolor{\PRCC}{T}he cross section \textcolor{\myModifColor}{starts to increase with the nuclearite mass due to the change in the cross section, resulting in more generated visible photons.}}
\label{fig:dndx}
\end{figure}

\textcolor{\PRCC}{A study has been performed in order to check if low speed particles such as nuclearities are able to trigger ANTARES filters}. In Fig. \ref{fig:dndx}, it is presented the expected number of visible photons per centimeter, as expected from Eq. \ref{eq:nbVisPhotons}, as a function of the nuclearite mass. Three positions at different levels \textcolor{\PRCC}{of the ANTARES detector} (on the top, middle and bottom of the detector string) are considered. In that figure, nuclearites are assumed \textcolor{\myModifColor}{to be} penetrating vertically to the detector with an initial velocity at the top of \textcolor{\PRCC}{the} atmosphere of $\beta\textcolor{\PRCC}{_{0}}=10^{-3}$. Nuclearites with a mass of $10^{13}$ GeV/c$^{2}$ generate less than $10^{3}$ photons per cm. \textcolor{\REVISION}{For reference, the number of Cherenkov photons in the sensitivity range of the PMTs emitted by a  muon is $\sim$  350/cm. However, the atmospheric muon events detected by ANTARES are mostly muon bundles which emit considerably more light per cm than single muons. For example, muon bundles with multiplicity of about 80 produce about 30000 photons/cm [21], and few events of this type are expected in the considered ANTARES livetime. Therefore, a mass of $4 \times 10^{13}$ GeV/c$^2$ is taken as a threshold mass in the analysis for nuclearite detection since it provides a light yield well separated from atmospheric events.}

When an event is triggered, all PMT pulses that happened within 2 $\mu$s before the first triggered hit and 2 $\mu$s after the last one are recorded. This collection of hits is referred to as a \textit{snapshot}. In the case in which two or more events have some overlapping hits, a merging of the events is done and a larger snapshot results. As nuclearites would typically have low velocities in the detector, both T3 and 3D triggers are expected to record a series of overlapped snapshots for these particles. 

% Fig. \ref{fig:trigEff} shows the efficiency of standard ANTARES triggers for nuclearites as a function of the number of L0 present in the event. The plot is produced for all simulated nuclearites masses and assuming isotropic arrival direction. 
\textcolor{\myModifColor}{
The combined efficiency of ANTARES triggers depends on the nuclearite mass and its arrival direction. When averaged over an isotropic flux from the upper hemisphere, the fraction of events with mass $>10^{14}$ GeV/c$^{2}$ recorded by the T3 or the 3D triggers ranges from 60\% to almost 100\% \textcolor{\PRCC}{depending on the mass}. The efficiency decreases for lower masses}. Table \ref{tab:tabEff} presents the overall efficiency for each nuclearite mass. It is defined as the number of triggered events divided by the number of generated events. The trigger efficiency increases from 25\% of generated events for nuclearites of $4 \times 10^{13}$ GeV/c$^2$ mass to about 97\% at the highest masses. The low trigger efficiency for low masses includes the effect of the energy loss of these particles from the generation surface of Fig. \ref{fig:simuHemi} before they reach the sensitive volume of the detector, and the fact that clipping particles are not luminous enough. These factors become negligible for high nuclearite masses.

%Only 25\% of generated events are triggered for $4 \times 10^{13}$ GeV/c$^2$ mass; in fact, nuclearites with lower masses would loss their energy before they reach the sensitive volume of the detector, as a result, small amount of events are triggered.

%Figure \ref{fig:trigEff} shows the trigger efficiency as function of nuclearites masses. The plot was produced for simulated nuclearites of all considred masses and assuming isotropic arrival direction. Only 25\% of events are triggered for the masses $4\times 10^{13}$ GeV/c$^2$, in fact, Low mass nuclearite events would lose their energy before reaching the sensitive volume of ANTARES which would result in low triggering efficiency as the detector would not be able to trigger such low energy events. 

% \begin{figure}[htbp]
% \centering
% \includegraphics[width=.8\textwidth, height=8cm]{figures/TEfficiency.pdf}

% \caption{The trigger efficiency for nuclearites as function of the number of L0 hits. See text for definition of T3 and 3D triggers.}
% \label{fig:trigEff}
% \end{figure}
 
\subsection{Preliminary quality cut on L0}
 
From the set of preselected runs, only events satisfying the predefined T3 or the 3D trigger conditions are included in this analysis. Two \textcolor{\myModifColor}{discriminating} variables are used to isolate nuclearites signals. The first one is a dimensionless variable denoted as \textcolor{\myModifColor}{log${_{10}(nhits3)/nfloor}$}, quantity that is proportional to the total amount of photons reaching the OMs of the detector. For a given event, $nhits3$ is the number of hits with charge of at least three photoelectrons, and $nfloor$ is the number of floors in the detector recording at least one hit for this particular event. The second discrimination variable is the event duration, $dt$, which represents the transit time of the event in the detector and refers to the time between the first and the last hit in the event. Nuclearites, as they are slow heavy particles, are expected to have a much larger transit time in the detector than background relativistic particles. \textcolor{\REV}{It is important to mention that by combining these variables, the efficiency of discriminating nuclearites events increases significantly compared to usage of a single variable, \textcolor{\MAU}{as for instance the number of large intensity hits (nhits3) or the number of involved floors (nfloor).}}
%, the distribution of $nfloor$ alone is shown from Fig. \ref{fig:nfloor}}

% \begin{figure}[htbp]
% \centering
% \includegraphics[width=.8\textwidth, height=8cm]{figures/nfloor.pdf}

% \caption{Distribution of nfloor variable for atmospheric muons and for considred nuclearites masses.}
% \label{fig:nfloor}
% \end{figure}

\begin{figure}[h]
\centering
\includegraphics[width=0.99\textwidth]{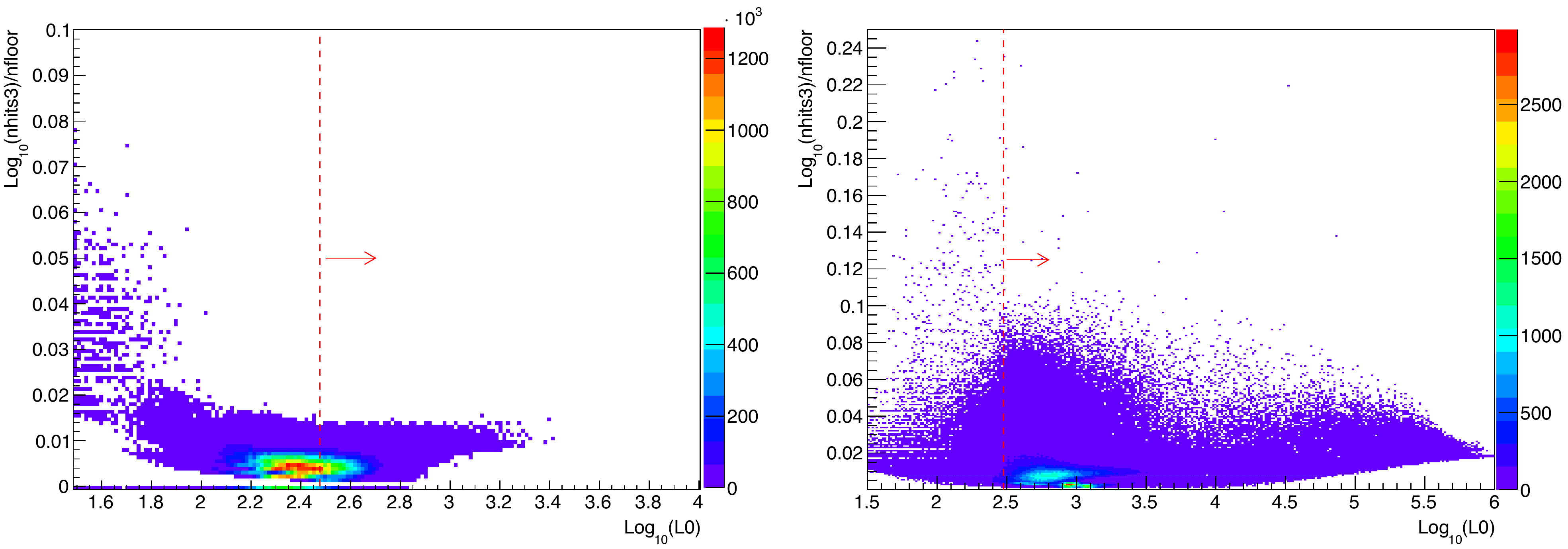}

\caption{Distribution of log$_{10}(nhits3)/nfloor$ versus log$_{10}(\text{L0})$ for simulated atmospheric muons \textcolor{\myModifColor}{(left), and for nuclearite\textcolor{\PRCC}{s} with a mass of $10^{16}$ GeV/c$^{2}$ (right)}. \textcolor{\myModifColor}{The palette of colors for the left panel represents the expected number of events in 839 days of livetime, while for \textcolor{\PRCC}{the} right panel represents the \textcolor{\LASTVR}{number} of simulated events}. The cut at 300 L0 hits (red dashed line) allows to reject \textcolor{\myModifColor}{background} events with high values of log$_{10}(nhits3)/nfloor$ characterized by a low number of L0 hits. Only 16\% of \textcolor{\PRCC}{the} simulated atmospheric muons survive the L0 cut.}
\label{fig:L0Cut}
\end{figure}

Due to their low speed and high masses, \textcolor{\REV}{nuclearite} events would generate a high number of visible photons around the detector in a relatively long period \textcolor{\myModifColor}{(reaching milliseconds}). They are expected to have high values for the log$_{10}(nhits3)/nfloor$ variable as $nfloor$ is \textcolor{\PRCC}{bound} by the total number of detector floors. However, %as shown in Fig. \ref{fig:nhnf} left, 
\textcolor{\PRCC}{atmospheric muon events with low number of L0 hits affect our signal} region in the log$_{10}(nhits3)/nfloor$ variable and they must be removed. \mbox {Fig. \ref{fig:L0Cut}} illustrates log$_{10}(nhits3)/nfloor$ versus log$_{10}$(L0) for simulated atmospheric muon events \textcolor{\PRCC}{(left)}  \textcolor{\PRCC}{and the same quantities are presented for nuclearite with a mass of $10^{16}$ GeV/c$^2$ (right).} 

 A \textcolor{\PRCC}{first} quality cut requiring at least 300 L0 hits for each event has been applied to clean our sample. This cut removes events firing with high amplitude the few PMTs present in a restricted region of the detector because of (for instance) a cascade of secondary particles induced by an atmospheric muon. Table \ref{tab:tabEff} \textcolor{\myModifColor}{summarises} the efficiency of the L0 cut for atmospheric muons and different masses of nuclearites. The efficiency is defined as the number of events after applying the L0 cut to the total number of  triggered events. The L0 cut removes \textcolor{\myModifColor}{85\%} of \textcolor{\PRCC}{the} simulated atmospheric muon events. 
%The efficiency is maximal for nuclearites with a mass of $10^{15}$ GeV/c$^{2}$. 

%\subsection{Discrimination variables}

%\begin{figure}[htbp]
%\centering
%\includegraphics[width=.7\textwidth, height=7cm]{L0effect09_17.pdf}
%
%\caption{Distribution of log$_{10}$(nhits3)/nfloor with and without the L0 cut for real data and simulated atmospheric muons.}
%\label{fig:L0effect}
%\end{figure}

\subsection{Selection efficiency}

\begin{figure}[h]
\centering
\includegraphics[width=.49\textwidth, height=5cm]{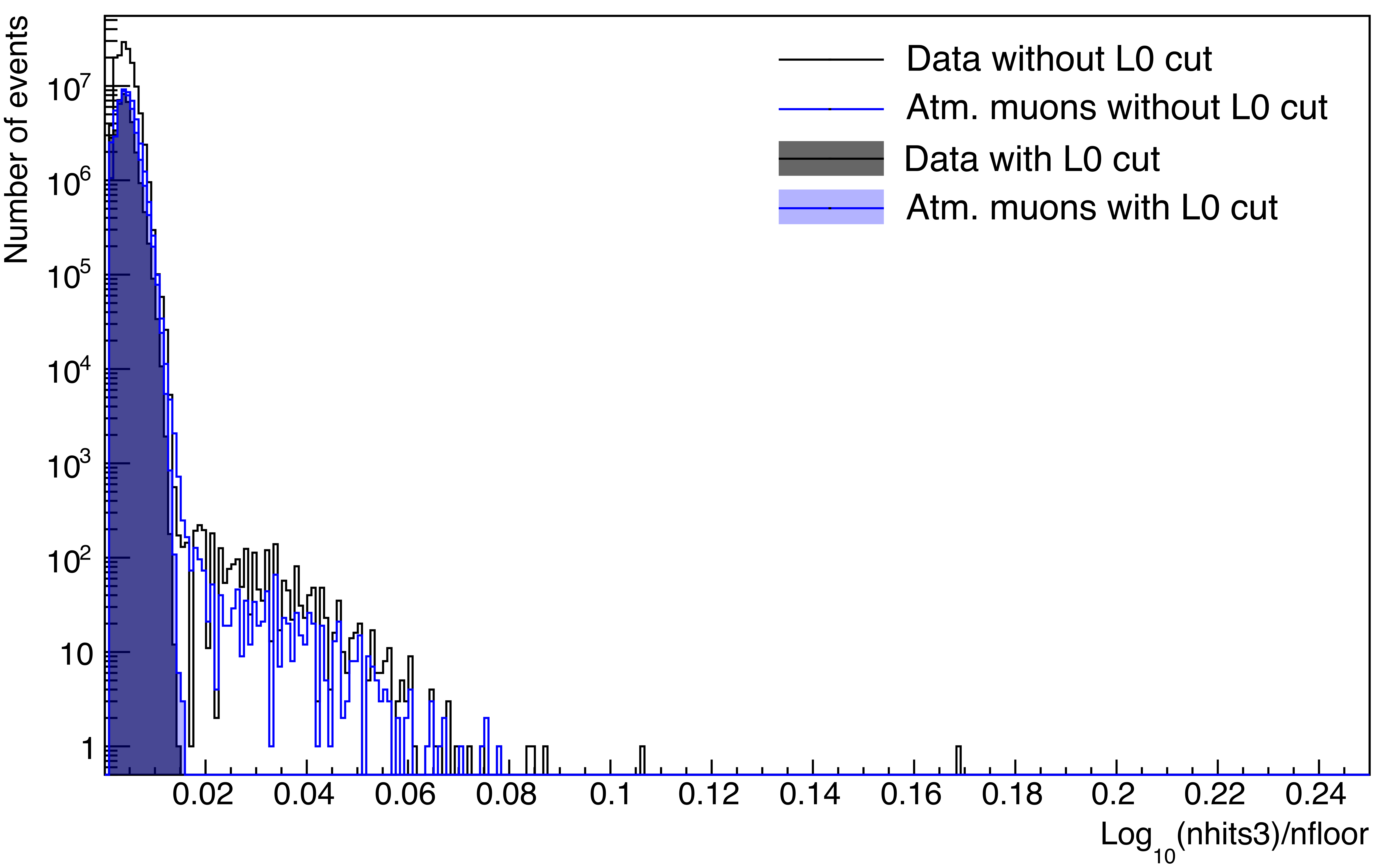}
\includegraphics[width=.49\textwidth, height=5cm]{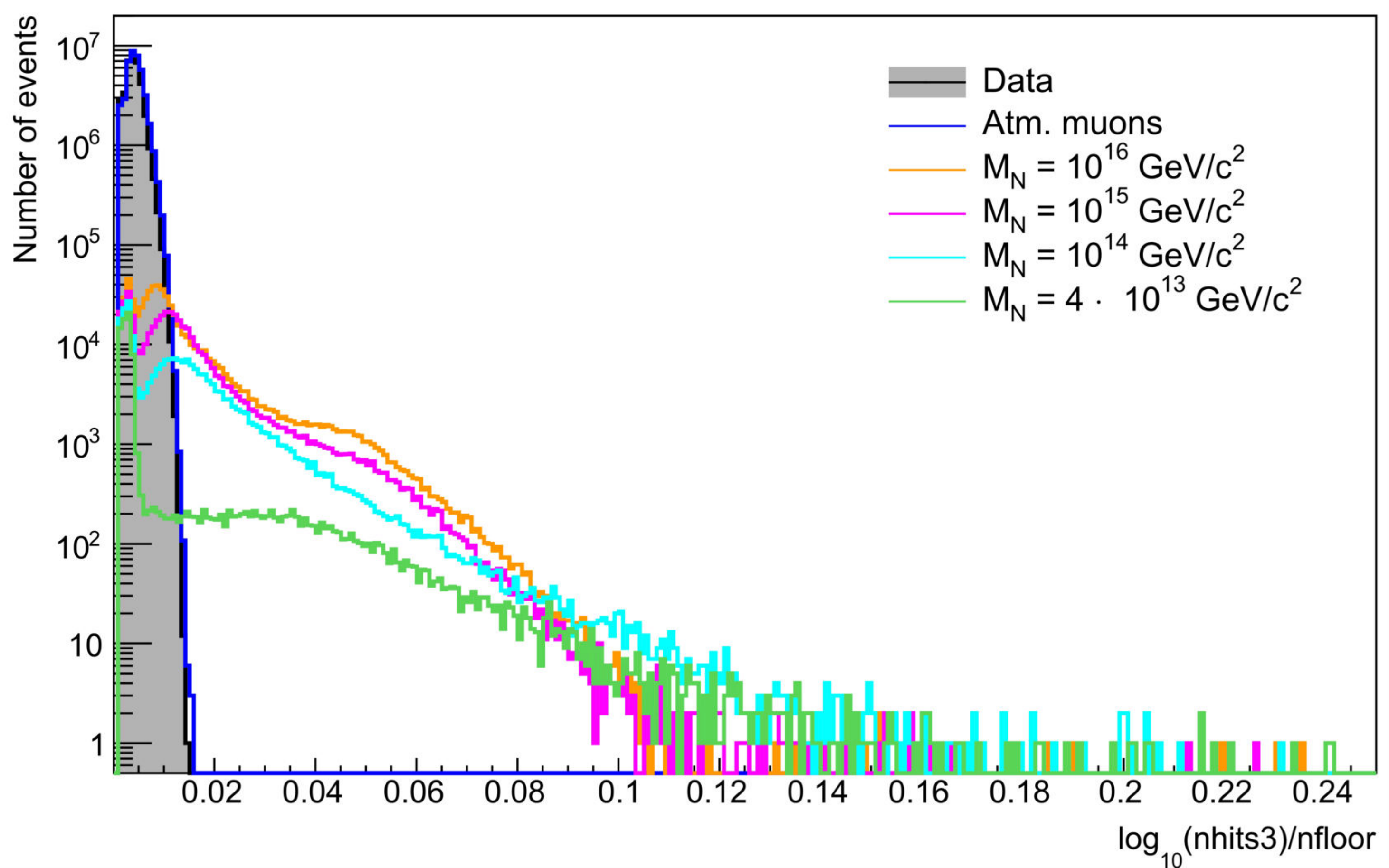}

\caption{Left: Distribution of the discrimination variable log$_{10}(nhits3)/nfloor$ without and with the cut L0 $\geq$ 300 on the total number of PMT hits using 839 days of livetime. The black histogram corresponds to data, while the blue one refers to simulated atmospheric muons. Right: The same distribution with the L0 $\geq$ 300 cut for data, simulated atmospheric muons and simulated and triggered nuclearites with different masses (other colors).}
\label{fig:nhnf}
\end{figure}

In order to discriminate nuclearite events from the background, variables that reflect the behavior of \textcolor{\myModifColor}{nuclearites} in the detector were used. \textcolor{\myModifColor}{Nuclearites} are expected to generate a high number of visible photons at the vicinity of the detector. This could result in a high number of fired floors with a high amplitude during a relatively long period.
%We define the nhits3 variable which is the number of hits with charge of at least 3 p.e. and nfloor variable which represent the number of floors that recorded at least one hit for a given event.
Fig. \ref{fig:nhnf} illustrates the log$_{10}(nhits3)/nfloor$ variable and shows the L0 cut effect on both data and atmospheric muons\textcolor{\PRCC}{,} for 839 days of livetime. From the right plot, most nuclearite events for different masses have a high value of this variable compared to muon events \textcolor{\PRCC}{whose distribution} stop at around \textcolor{\PRCC}{$\sim$$0.015$}. The signal of these particles is also \textcolor{\myModifColor}{characterised} by a long snapshot duration.

\begin{figure}[htbp]
\centering
\includegraphics[width=1\textwidth]{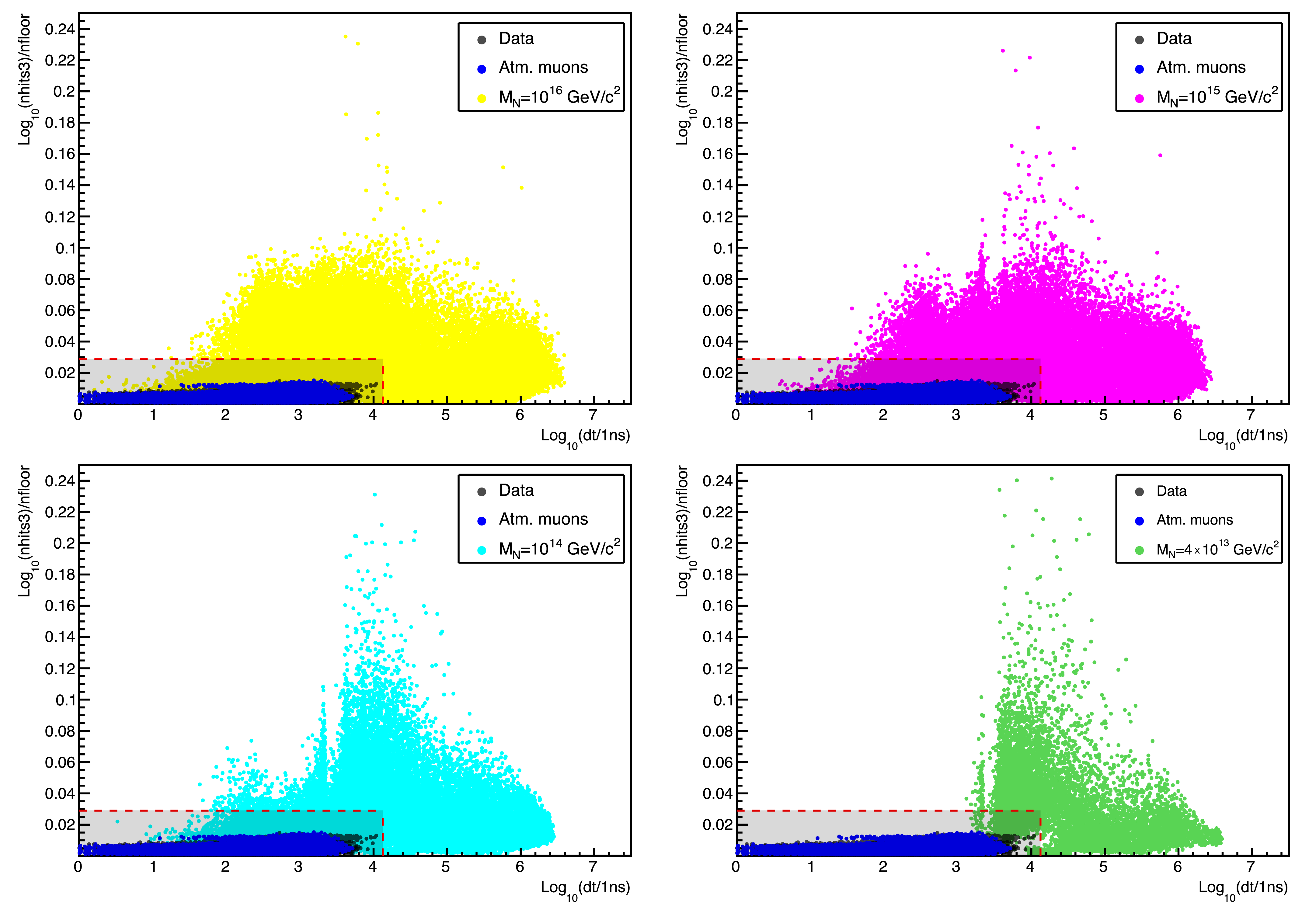}

\caption{Scatter plot of \textcolor{\PRCC}{log$_{10}(nhits3)/nfloor$} versus log$_{10}(dt \textcolor{\myModifColor}{\text{/1ns}})$ for events with at least 300 L0 hits. Real data in black, atmospheric muons in blue, and for different nuclearite masses. The red dotted lines represent the optimised cuts on both discrimination variables.}
\label{fig:nhnh_dt}
\end{figure}

In order to reduce the background, events are selected based on both parameters characterizing the signal. Fig. \ref{fig:nhnh_dt} shows log$_{10}(nhits3)/nfloor$ versus $dt$ for each nuclearite mass. \textcolor{\myModifColor}{An important fraction of} nuclearite events show a long duration, $dt$, typically higher than $3 \times 10^{3}$ ns, combined with high values of log$_{10}(nhits3)/nfloor$. The scatter plots in Fig. \ref{fig:nhnh_dt} show that both variables are well suited to discriminate between nuclearites and background from cosmic ray muons. By applying the appropriate cuts, the region dominated by the noise could easily be discriminated from the region relevant for our signal. However, in order to \textcolor{\myModifColor}{maximise} the performance of these cuts, an optimisation is required. The methods used to \textcolor{\myModifColor}{optimise} the cuts \textcolor{\MAU}{are discussed in the next subsection. As a result, the signal to background discrimination is maximized by removing events according to the conditions:
}

\begin{align}
   \text{log}_{10}(dt/1 \text{ns}) < 4.125 \ .AND. \ \text{log}_{10}(nhits3)/nfloor < 0.025 
\end{align}

Table \ref{tab:tabEff} shows the selection efficiency of the \textcolor{\myModifColor}{optimised} cuts for each nuclearite mass. 
%The selection  efficiency decreases when the mass increases since most high mass nuclearite events are overlapped to background in both variables.

\begin{table}[h]
\centering
\begin{tabular}{|l|c|c|c|c||c|}
\hline 
Nuclearite mass (GeV/c$^2$)& $10^{16}$ & $10^{15}$ & $10^{14}$ & $4\times 10^{13}$ & Atm. muons\\ 
\hline 
Trigger efficiency (\%) & 97 & 88 & 61 & 25 & 9.5 \\ 
\hline 
L0 cut efficiency (\%) & 40 & 42 & 31 & 17 & 15 \\ 
\hline 
Selection efficiency (\%) & \textcolor{\PRCC}{18} & \textcolor{\PRCC}{19} & \textcolor{\PRCC}{25} & \textcolor{\PRCC}{68} & 0 \\ 
\hline \hline 
Final efficiency (\%) & \textcolor{\PRCC}{6.9} & \textcolor{\PRCC}{7.0} & \textcolor{\PRCC}{4.7} & \textcolor{\PRCC}{2.8} & 0 \\ 
\hline 

\end{tabular} 

\caption{Selection efficiencies as a function of the nuclearite masses at the different steps discussed in the text. The trigger efficiency in the first row is \textcolor{\myModifColor}{computed with} respect to the number of simulated events, while the \textcolor{\myModifColor}{efficiencies} in the following rows \textcolor{\myModifColor}{are computed with respect to} the previous one. \textcolor{\myModifColor}{The selection efficiency of atmospheric muons are also reported in the last column.}}
\label{tab:tabEff}
\end{table}

%\subsection{The rejection factor}

%From figure \ref{fig:nhnf}, a good agreement between experimental data and MC simulation is observed. Most nuclearites events for different mass have a high number of log$_{10}$(nhits3)/nfloor compared to muons events which stops to $\sim 0.015$. If we select events above this value, we can isolate our signal. %However, in order to maximize the signal and so, the response of ANTARES to these particles, we use the second variable $dt$ to constraint the selected region by cuts on these two isolation variables. The second cut allows to gain more signal events, namely, those with log$_{10}$(dt) $>$ 4, taking into account the cut on the first variable.

\subsection{Optimization of the two variables}

In order to achieve the best sensitivity of ANTARES to nuclearites without any bias, %therefore, the best flux upper limit if no events are found, 
the Model Rejection Factor (MRF) is used \cite{Hill} to \textcolor{\myModifColor}{optimise} the cuts on the discrimination variables taking into account the statistical fluctuations in the simulated atmospheric muons distributions. The \textcolor{\PRCC}{MRF} technique consists of varying the cuts with small steps until the minimum of MRF is reached, this minimum corresponds to the values that give the best sensitivity.

The sensitivity at 90\% confidence level (CL), denoted as $S_{90}$, is computed using the Feldman-Cousins method \cite{FeldmanCousins}, assuming events with a Poissonian distribution:
\textcolor{\PRCC}{
\begin{equation}
\bar{\mu}_{\text{90}} =  \sum_{n_{\text{obs}}=0}^{\infty} \mu_{\text{90}}(n_{\text{obs}}, n_{\text{b}}) \times \dfrac{n_{\text{b}}^{n_{\text{obs}}}}{n_{\text{obs}} !} \times e^{-n_{\text{b}}},
\label{eq:mu90}
\end{equation}
}
\textcolor{\PRCC}{
\begin{equation}
S_{\text{eff}} = \dfrac{n_{\text{Nuc}}}{\Phi_{\text{Nuc}}},
\label{eq:sEff}
\end{equation}
}
\textcolor{\PRCC}{
\begin{equation}
S_{90} =\dfrac{\bar{\mu}_{90}(n_{\text{b}})}{S_{\text{eff}} \times T},
\label{equ:sensEqua}
\end{equation}
}
\noindent where $n_{\text{obs}}$ is \textcolor{\PRCC}{the} number of observed events and $n_{\text{b}}$ the number of expected background events from \textcolor{\PRCC}{the full} dataset. $T$ is the duration of data taking corresponding to the 2009-2017 period \textcolor{\myModifColor}{and $\mu_{\text{90}}(n_{\text{obs}}, n_{\text{b}})$ represents the upper limit for $n_{\text{obs}}$ events and $n_{\text{b}}$ expected backgroung events}. $n_{\text{Nuc}}$ \textcolor{\PRCC}{denotes} the number of nuclearites remaining after applying the optimized cuts, and $\Phi_{\text{Nuc}}$ the \textcolor{\PRCC}{corresponding} flux of generated nuclearites.

\begin{figure}[ht]
     \centering
     \includegraphics[width=.8\textwidth, height=8cm]{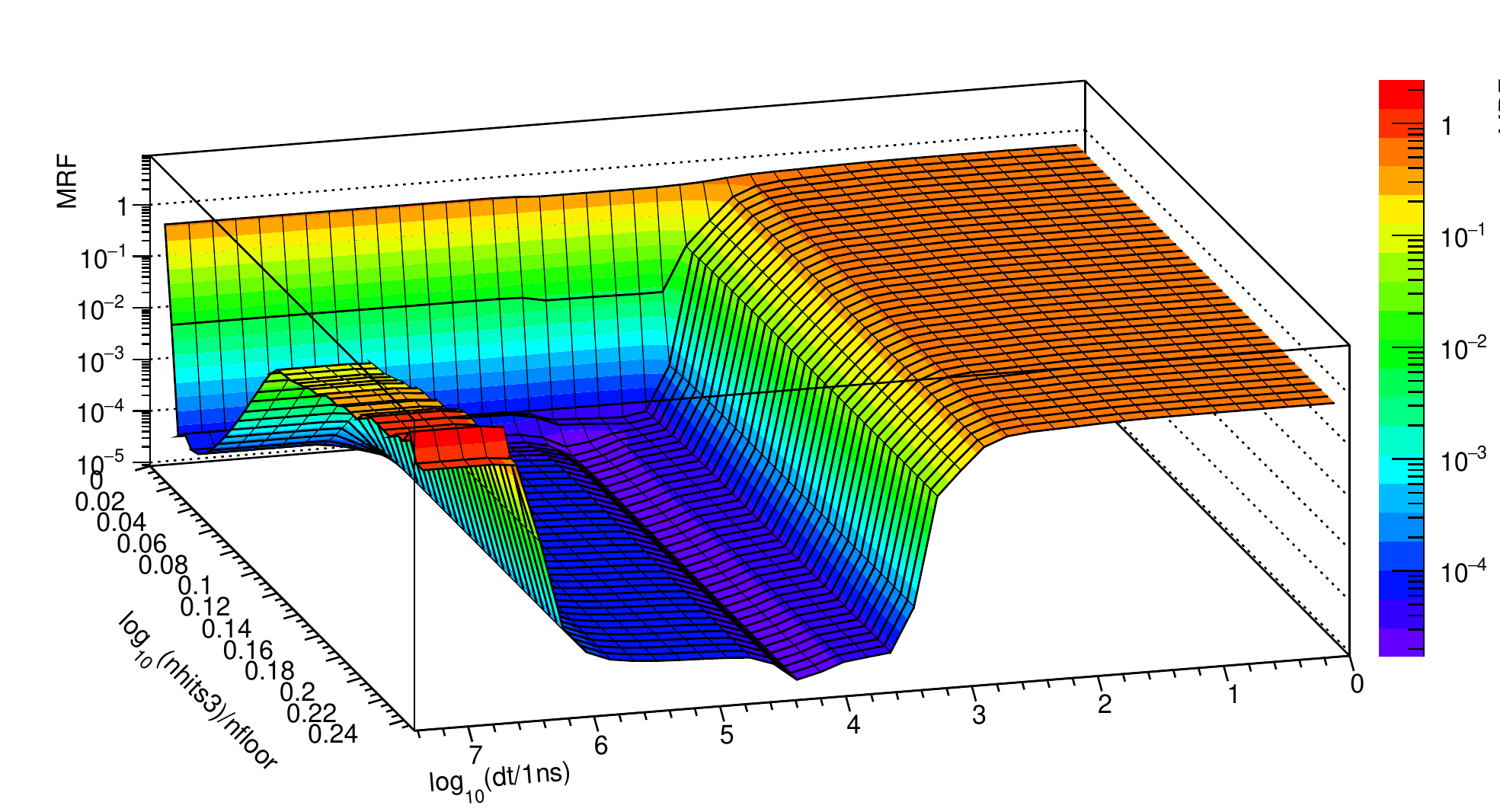}
    
     \caption{MRF as function of log$_{10}(nhits3)/nfloor$ and $dt$ for the highest nuclearite mass \textcolor{\PRCC}{considered} ($10^{16}$ GeV/c$^2$).}
     \label{fig:MRF}
     \centering
\end{figure}

The best values on $dt$ and log$_{10}(nhits3)/nfloor$ are those minimising the MRF for each nuclearite mass. %Nevertheless, both distributions tails are suffering from a lack of statistics in the signal region, and it might affect our results. To recover for this, we have made an extrapolation in these regions based on a fit with a Landau type function. The optimization as well as the final results are taking into account this extrapolation. 
 Fig. \ref{fig:MRF} shows an example of the MRF for the \textcolor{\myModifColor}{highest} nuclearite mass as a function of the selection parameters. The values that minimise the MRF for the four values of the \textcolor{\REV}{simulated} nuclearite masses are reported in the first two rows of Table \ref{tab:values}. The \textcolor{\PRCC}{fraction of event} passing all the selection criteria is reported in the third row, while the number of remaining background events after the selection cuts and the value of the MRF are in the rows 4 and 5, respectively.
 %The best values are those where the MRF is minimal.

%%%%%%%%%%%%%%%%%%%%%%%%%%%%%%%%%%%%%%%%%%%%%%%%%%%%%%%%%

\if 0
\begin{figure}[H]
\centering
\includegraphics[width=.7\textwidth, height=7cm]{Fit_Extrapolation_nhnf_0data.png}

\caption{Distribution of log$_{10}$(nhits3)/nfloor with the L0 cut for experimental data. The distribution tail is dominated by optical background and present statistical fluctuation. The fit (in red) has been made using a Landau type function, based on the fit parameters, an extrapolation has been made in the signal region (in blue). }
\label{fig:extranhnf}
\end{figure}

\begin{figure}[h]
\centering
\includegraphics[width=.7\textwidth, height=7cm]{dt_extrap_l0cut300_09_17_data.png}

\caption{Distribution of log$_{10}(dt)$ for real data for events with at least 300 L0 hits. The first pique at around 200 ns is due to very fast relativistic particles, the second one is mainly due to the events triggered by the 3D trigger, while the sharp cut at around 4000 ns is due to maximum time window allowed in the T3 trigger. The fit is done using a Landau type function.}
\label{fig:extradt}
\end{figure}
\fi

%% file: sections/resultsAndDiscusion.tex
\textcolor{\MAU}{
\section{ Systematic effects and experimental results}
}
\label{sec:resultsAndDiscussion}
\textcolor{\MAU}{
\subsection{Systematic uncertainties}
\label{sec:sysErr}
}
\textcolor{\MAU}{
Experimental searches of nuclearites rely on the crossing time of the particles in the detector and the light yield per unit length of the track, as given in the \mbox{De R\'{u}jula} and Glashow paper \cite{DeRujulaGlashow}. Following \cite{DeRujulaGlashow}, in the present paper the luminous efficiency $\eta$ as derived for transparent water in the [550-330] nm wavelength range was used. This wavelength region is well within sensitivity of ANTARES PMTs. The energy distribution of photons follows the black-body distribution, which for nuclearites arriving in the detector with masses above the threshold of \mbox{$4\times10^{13}$ GeV/c$^2$} has a peak at wavelengths shorted than 300 nm. The intensity of Cherenkov emission produced by relativistic particles is proportional to $1/\lambda^2$, and also in this case short wavelengths dominate. When the two different spectra are folded in our simulations with the water transmission \cite{ref133} and the quantum efficiency the PMTs as a function of the wavelength \cite{ref132}, the differences between the two different spectra in the photoelectron yields on the PMTs are within 30-40\%. The minimum nuclearite mass ($4\times 10^{13}$ GeV/c$^2$) considered in the analysis ensures that the number of emitted photons per cm are of the order of 10$^6$ at each detector depth: as a consequence, the trigger and selection efficiencies are not affected by the variation in the emission spectra of visible photons.} 

\textcolor{\MAU}{
The uncertainties for the signal produced by atmospheric muons are derived from the statistical fluctuations in the signal region for both discrimination variables ($d$t and log$_{10}(nhits3)/nfloor$); this is recovered by extrapolating the distributions using a Landau type function (see ref. \cite{Jihad} for further details). Systematic uncertainties on the effect of atmospheric muon events are mainly due to the detector description and to the knowledge on the environmental parameters. These are mainly related to the uncertainties on the angular acceptance of the optical modules and on the light absorption and scattering lengths in the sea water \cite{ref132,ref133}. Hence, an overall of +35\% -30\% effect on the expected atmospheric muon rate results from $\pm$15\% as a maximum of uncertainty on the optical module acceptance and $\pm$10\% on the light absorption in water along the entire wavelength spectrum \cite{systematics}. \textcolor{\REVISION}{ The computation of the number of observed events $n_{obs}$ and the number of expected background events $n_b$ given in equation \ref{eq:mu90} incorporate systematic uncertainties using the method described in \cite{conrad}.
}
}
% The sensitivity at 90\% confidence level (C.L.), noted $S_{90}$, is computed using the Feldman-Cousins formula \cite{FeldmanCousins}, assuming events with a Poissonian distribution:

% \begin{equation}
% S_{90} =\dfrac{\bar{\mu}_{90}(n_{b})}{S_{eff} \times T},
% \label{equ:sensEqua}
% \end{equation}

% \begin{equation}
% \bar{\mu}_{90} =  \sum_{n_{obs}=1}^{\infty} \mu_{90}(n_{obs}, n_{b}) \times \dfrac{n_{b}^{n_{obs}}}{n_{obs} !} \times e^{-n_{b}},
% \end{equation}

% \begin{equation}
% S_{eff} = \dfrac{n_{Nuc}}{\Phi_{Nuc}},
% \end{equation}

% \noindent where, $n_{obs}$ is number of observed events and $n_{b}$ the number of expected background events from all the dataset. T is the duration of data taking corresponding to the 2009-2017 period in seconds, $n_{Nuc}$ represents the number of nuclearites remaining after applying the optimized cuts, and $\Phi_{Nuc}$ represents the flux of generated nuclearites.

%The effect of the cuts (reported in Table \ref{tab:values}) on simulated signal and background events is to keep \textcolor{\myModifColor}{most of the} backgroung events in the \textcolor{\myModifColor}{third} quadrant (highlited in Fig. \ref{fig:nhnh_dt}), whose axes are defined by the values of the reported cut. The optimisation procedure finds the same value for each parameter in the four simulated masses. 

\textcolor{\MAU}{
\subsection{Unblinding results}
}
After unblinding the remaining 90\% of data not used for testing the optimisation procedure, no event is present in the region \textcolor{\PRCC}{beyond the optimised cuts}. %with log$_{10}(nhits3)/nfloor \geq$ 0.029 and log$_{10}$(dt) $\geq$ 4.125. 
Since no event \textcolor{\PRCC}{is} found,  the upper limit on the flux of nuclearites at 90\% CL is computed as

\begin{equation}
\phi_{\text{90}} =\dfrac{{\mu}_{\text{90}}(n_{\text{obs}}, n_{\text{b}})}{S_{\text{eff}}\times T},
\label{fluxFormula}
\end{equation}

\noindent
where the confidence interval at 90\% CL $\mu_{90}(n_{\text{obs}}, n_{\text{b}})$ is computed from the unified approach of Feldman-Cousins \cite{FeldmanCousins}. \textcolor{\REVISION}{$S_{eff}$ is defined in equation \ref{eq:sEff} and $T$ is the livetime of the analysis.}
%$S_{\text{eff}}$ and $T$ are defined in Eq. \ref{equ:sensEqua}.

\begin{table}[]
\centering
\resizebox{\textwidth}{!}{
\begin{tabular}{|l|c|c|c|c|}
\hline 
 Nuclearite mass (GeV/c$^2$)
 & $10^{16}$  & $10^{15}$ & $10^{14}$ & $4\times 10^{13}$ \\ 
\hline 
Best cut on log$_{10}$(\textcolor{\PRCC}{\textit{dt}}\textcolor{\myModifColor}{/1ns}) & \textcolor{\PRCC}{$\geq$ 4.27} & \textcolor{\PRCC}{$\geq$ 4.27} & \textcolor{\PRCC}{$\geq$ 4.27} & \textcolor{\PRCC}{$\geq$ 4.35} \\ 

Best cut on \textcolor{\PRCC}{log$_{10}(nhits3)/nfloor$} &  \textcolor{\PRCC}{$\geq$ 0.029} & \textcolor{\PRCC}{$\geq$ 0.024} & \textcolor{\PRCC}{$\geq$ 0.024} & \textcolor{\PRCC}{$\geq$ 0.024} \\ 

Remaining nuclearites (\%) & \textcolor{\PRCC}{6.9} & \textcolor{\PRCC}{7.0} & \textcolor{\PRCC}{4.7} & \textcolor{\PRCC}{2.8} \\ 

Remaining background  & \textcolor{\PRCC}{1.05} & \textcolor{\PRCC}{0.35} & \textcolor{\PRCC}{0.35} & \textcolor{\PRCC}{0.31} \\ 

MRF & \textcolor{\PRCC}{1.6$\times 10^{-5}$} & \textcolor{\PRCC}{2$\times 10^{-5}$} & \textcolor{\PRCC}{2.7$\times 10^{-5}$} & \textcolor{\PRCC}{4.12$\times 10^{-5}$}  \\ 

Flux upper limit 90 \%CL (cm$^{-2} \times$ sr$^{-1} \times$ s$^{-1}$) & \textcolor{\PRCC}{6.6$\times 10^{-18}$} & \textcolor{\PRCC}{8.1$\times 10^{-17}$} & \textcolor{\PRCC}{1.1$\times 10^{-17}$} & \textcolor{\PRCC}{1.6$\times 10^{-17}$} \\ 
\hline 
\end{tabular} 
}
\caption{Values of the optimised parameters and the results of the MRF and \textcolor{\myModifColor}{upper limits} for the different nuclearite masses for 839 days of livetime.}
\label{tab:values}
\end{table}

%Table \ref{tab:values} summarises the optimized parameters as well as the remaining events for nuclearites and background. No event has survived beyond the cuts from the real data, the upper limit on the flux of nuclearites at 90\% C.L. for nine years of data taking with ANTARES may be computed as follows:

The 90\% CL flux upper limit values reported in Table \ref{tab:values} include both statistical and systematic uncertainties. 
%The statistical uncertainties for atmospheric muons are derived from the statistical fluctuations in the signal region for both discrimination variables; this is recovered by extrapolating the distributions using a Landau type function. Systematic uncertainties for atmospheric muons \textcolor{\PRCC}{background events} are mainly due to the detector description and to the knowledge on the environmental  parameters. 
%Systematic errors are mainly related to the angular acceptance of the optical modules and to the light absorption and scattering lengths in the sea water \cite{ref132,ref133}. Hence, an overal of $^{+35\%}_{-30\%}$ effect on the expected \textcolor{\PRCC}{atmospheric} muon rate results from $\pm$15\% as a maximum of uncertainty on the optical module acceptance and $\pm$10\% on the light absorption in water along the entire wavelength spectrum \cite{systematics}.

\begin{figure}[h]
     \centering
     \includegraphics[width=.8\textwidth, height=8cm]{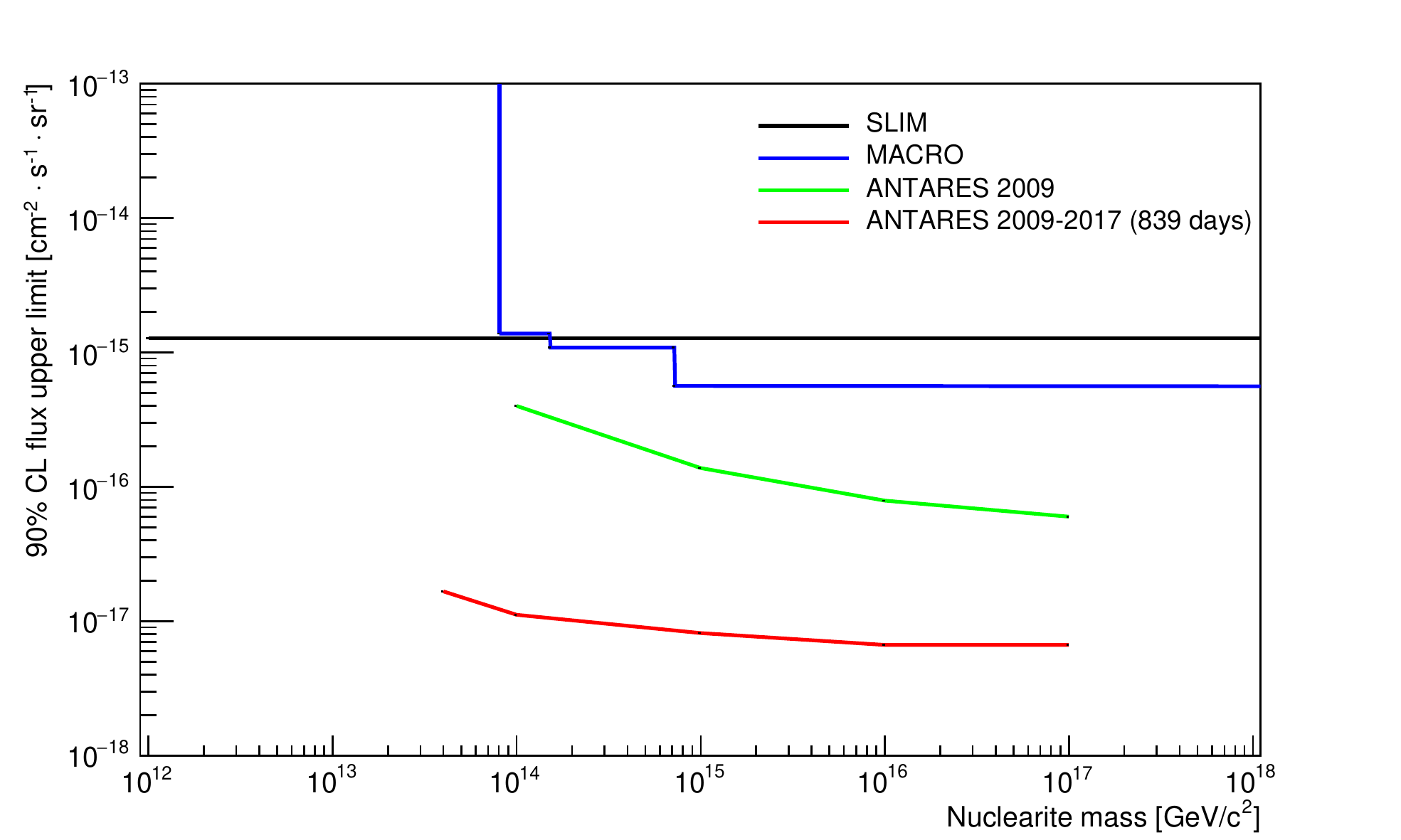}
    
     \caption{Upper limit on the flux of nuclearites \textcolor{\myModifColor}{with Galactic velocities ($\beta = 10^{-3}$) in} red line, using 839 days of livetime in the period 2009-2017 with the ANTARES detector. The green line corresponds to a \textcolor{\PRCC}{previous} ANTARES result obtained with \textcolor{\myModifColor}{a smaller} data sample \cite{Pavalas}. Results from other experiments, MACRO \cite{MACRO} and SLIM \cite{SLIM}, are also shown.}
     \label{fig:flux}
\end{figure}

The obtained upper limit (UL) on the nuclearite flux for $T=839$ days of livetime of the \textcolor{\myModifColor}{ANTARES detector} is shown in Fig.  \ref{fig:flux}. %Therefore, ANTARES is sensitive to slow moving heavy particles such as cosmic nuclearites. 
For masses higher than $10^{16}$  GeV/c$^{2}$, nuclearites  emit more light. Therefore, the limit of the last test point can be taken as a conservative limit also for higher nuclearite masses, up to the value of $\sim 10^{17}$  GeV/c$^{2}$ where detector saturation effects \mbox{start to occur}.

%% file: sections/conclusion.tex
\vspace{0.5cm}
\section{Conclusion}
\label{sec:conclusion}

A search for nuclearites in the mass range from $4 \times 10^{13}$  GeV/c$^{2}$ to $10^{16}$  GeV/c$^{2}$ reaching the ANTARES neutrino telescope using nine years of data (for an equivalent livetime of 839 days) is presented. Typical Galactic velocities ($\beta = 10^{-3}$) at the surface of the Earth atmosphere have been considered for this study. After propagation down to the sea level, a large fraction of events trigger the apparatus and \textcolor{\myModifColor}{they are} discriminated against the background \textcolor{\myModifColor}{formed by} atmospheric muon bundles. The selection efficiency increases with increasing nuclearite masses. No events survive the selection cuts and 90\% CL upper limits are derived as a function of the nuclearite mass. \textcolor{\myModifColor}{These are the most stringent limits ever set for nuclearites \textcolor{\PRCC}{with Galactic velocities} and the first ones set by a neutrino telescope}.

\if false

 we presented the results obtained for the search of down-going heavy and slow moving nuclearites in the ANTARES ANATRES detector

This work, shows the  assuming a velocity at the entrance of the atmosphere of $\beta_{0}=10^{-3}$. Nuclearites heavier than $4 \cdot 10^{13} $ GeV/c$^2$ would be able to reach the  depth with enough energies to generate a sufficient visible photon and to be detected by the PMTs. The analysis was very sensitive to the optical background characterizing the detection medium. Hence, restrictive selection conditions were applied to each run to make sure of its eligibility. Moreover, each event should satisfy a set of preselection conditions to be admitted, this proceeder allow to choice a clean data even in the presence of the bio-optical noise, namely, bioluminescence and $^{40}$K decay which presents a background for the study.

where atmospheric muons and optical background. Before the unblinding of the data, a burned sample (runs ending with 0, corresponding to 10\% of the data) of real data in the period of 2009-2017 is used to compare data-MC and to optimize the cuts on the discrimination variables, and finally to compute the sensitivity for a given mass of nuclearites for the period in consideration. The usage of the model of rejection factor allowed us to calculate the optimized cuts on $dt$ and log$_{10}$(nhits3)/nfloor takin into account the data extrapolated events which make the cuts safer from the optical noise, and so, calculate the best sensitivity or limit of ANTARES to these particles.

The upper limit obtained is improving the limits reported by other experiments such as MACRO and SLIM. It also improves the results already obtained by ANTARES for 2009 period. 
 \fi

%% file: other/Ackn.tex
\section*{Acknowledgments}
The authors acknowledge the financial support of the funding agencies:
% France:
Centre National de la Recherche Scientifique (CNRS), Commissariat \`a
l'\'ener\-gie atomique et aux \'energies alternatives (CEA),
Commission Europ\'eenne (FEDER fund and Marie Curie Program),
Institut Universitaire de France (IUF), LabEx UnivEarthS (ANR-10-LABX-0023 and ANR-18-IDEX-0001),
R\'egion \^Ile-de-France (DIM-ACAV), R\'egion
Alsace (contrat CPER), R\'egion Provence-Alpes-C\^ote d'Azur,
D\'e\-par\-tement du Var and Ville de La
Seyne-sur-Mer, France;
% Germany: 
Bundesministerium f\"ur Bildung und Forschung
(BMBF), Germany; 
% Italy
Istituto Nazionale di Fisica Nucleare (INFN), Italy;
% Netherlands
Nederlandse organisatie voor Wetenschappelijk Onderzoek (NWO), the Netherlands;
% Romania
Executive Unit for Financing Higher Education, Research, Development and Innovation (UEFISCDI), Romania;
% Spain
Ministerio de Ciencia, Innovaci\'{o}n, Investigaci\'{o}n y
Universidades (MCIU): Programa Estatal de Generaci\'{o}n de
Conocimiento (refs. PGC2018-096663-B-C41, -A-C42, -B-C43, -B-C44
and refs. PID2021-124591NB-C41, -C42, -C43)
(MCIU/FEDER), Generalitat Valenciana: Prometeo (PROMETEO/2020/019),
Grisol\'{i}a (refs. GRISOLIA/2018/119, /2021/192) and GenT
(refs. CIDEGENT/2018/034, /2019/043, /2020/049, /2021/023) programs, Junta de
Andaluc\'{i}a (ref. A-FQM-053-UGR18), La Caixa Foundation (ref. LCF/BQ/IN17/11620019), EU: MSC program (ref. 101025085), Spain;
% Marocco
Ministry of Higher Education, Scientific Research and Innovation, Morocco, and the Arab Fund for Economic and Social Development, Kuwait.
% A.O.B.:
We also acknowledge the technical support of Ifremer, AIM and Foselev Marine
for the sea operation and the CC-IN2P3 for the computing facilities.

%% file: article.bbl
\begin{thebibliography}{9}

\bibitem{Bodmer}  %1
A. Bodmer, 
\textit{Collapsed Nuclei}, 
\href{https://doi.org/10.1103/PhysRevD.4.1601}{Phys. Rev. D 4 (1971) 1601}.

\label{bib:Bodmer}

%1
\bibitem{EWitten}    %3
E. Witten, 
\textit{Cosmic separation of phases},
\href{https://doi.org/10.1103/PhysRevD.30.272}{Phys. Rev. D 30 (1984) 272-285}.
\label{bib:EWitten}


\bibitem{Terazawa}   %2
H. Terazawa,
\textit{Super-Hypernuclei in the Quark-Shell Model. II
},
\href{https://doi.org/10.1143/JPSJ.58.4388}{J. Phys. Soc. Japan, 58 (1989), 4388-4393}.
\label{bib:Terazawa}

\bibitem{EFarhiRLJaffe} %4
E. Farhi and R. L. Jaffe,
\textit{Strange matter},
\href{https://doi.org/10.1103/PhysRevD.30.2379}{Phys. Rev. D 30 (1984), 2379}.
\label{bib:EFarhiRLJaffe}

\bibitem{Bauswein} % 5   % general SQM definition
A. Bauswein, H.-T. Janka, R. Oechslin, G. Pagliara, I. Sagert, J. Schaffner-Bielich, M. M. Hohle, and R. Neuhäuser,
\textit{Mass Ejection by Strange Star Mergers and Observational Implications},
\href{https://doi.org/10.1103/PhysRevLett.103.011101}{Phys. Rev. Lett. 103 (2009), 011101}.
\label{bib:Bauswein}


\bibitem{Chodos} %  6 % MIT bag model
A. Chodos, R. L. Jaffe, K. Johnson, and C. B. Thorn, 
\textit{Baryon structure in the bag theory},
\href{https://doi.org/10.1103/PhysRevD.10.2599}{Phys. Rev. D 10 ( 1974), 2599}.
\label{bib:Chodos}

\bibitem{DeGrand} %  7  % MIT bag model
T. DeGrand, R. L. Jaffe, K. Johnson, and J. Kiskis,
\textit{Masses and other parameters of the light hadrons},
\href{https://doi.org/10.1103/PhysRevD.12.2060}{Phys. Rev. D 12 (1975), 2060}.
\label{bib:DeGrand}

\bibitem{Greiner}   %  8  % general SQM definition
C. Greiner, A. Diener, J. Schaffner, H\textcolor{\PRCC}{.} Stoecker,
\textit{Strange matter — a new domain of nuclear physics},
\href{https://doi.org/10.1016/0375-9474(94)90620-3}{Nucl. Phys. A 566 (1994), 157-165}.
\label{bib:Greiner}

\bibitem{Bjorken}  % 9 % SQM in cosmic rays
J. D. Bjorken and L. D. McLerran,
\textit{Explosive quark matter and the "Centauro" event},
\href{https://doi.org/10.1103/PhysRevD.20.2353}{Phys. Rev. D 20 (1979), 2353}.
\label{bib:Bjorken}

\bibitem{DeRujulaGlashow} 
A. De R\'{u}jula and S. L. Glashow, 
\textit{Nuclearites - a novel form of cosmic radiation}, 
\href{https://doi.org/10.1038/312734a0}{Nature 312 (1984), 734–737}.
\label{bib:DeRujulaGlashow}

\bibitem{Madsen}    % general SQM definition
J. Madsen,
\textit{Strangelet propagation and cosmic ray flux},
\href{https://doi.org/10.1103/PhysRevD.71.014026}{Phys. Rev. D 71 (2005), 014026}.
\label{bib:Madsen}

\bibitem{Maurizio}    % general SQM definition
M. Spurio,
\textit{Searches for Magnetic Monopoles and other Stable Massive Particles},
\href{https://arxiv.org/abs/1906.02039}{arXiv: 1906.02039 [hep-ph]}.
\label{bib:Maurizio}

\bibitem{Pavalas}
G. E. P\u{a}v\u{a}la\c{s} [ANTARES Collaboration],
\textit{Search for nuclearites with the ANTARES neutrino telescope},
\href{https://doi.org/10.22323/1.236.1060}{Proceedings of The 34th International Cosmic Ray Conference {\textemdash} PoS(ICRC2015), 236 (2015), 1060}.
\label{bib:Pavalas}

%4

\bibitem{ARS}
\textcolor{\myModifColor}{J. A. Aguilar} [ANTARES Collaboration], 
\textit{Performance of the front-end electronics of the ANTARES neutrino telescope}, 
\href{https://doi.org/10.1016/j.nima.2010.06.225}{Nucl. Instrum. Meth. A 622 (2010) 59 [arXiv:1007.2549]}.
\label{bib:ARS}

\bibitem{Aguilar}
M.~Ageron [ANTARES Collaboration],
\textit{ANTARES: the first undersea neutrino telescope}, 
\href{https://doi.org/10.1016/j.nima.2011.06.103}{Nucl. Instrum. Meth. A 656 (2011), Issue 1, 11-38}.
\label{bib:Aguilar}

\bibitem{Chin}
S. A. Chin and A. K. Kerman,
\textit{Possible Long-Lived Hyperstrange Multiquark Droplets}, 
\href{https://doi.org/10.1103/PhysRevLett.43.1292}{Phys. Rev. Lett. 43, 1292}.
\label{bib:Chin}

\bibitem{Shibata}
T. Shibata, 
\textit{Study of Shower Phenomena in the Atmosphere. I: Analytical Derivation of Lateral Structure Function of Hadronic Components},
\href{https://doi.org/10.1143/PTP.57.882}{Prog. Theor. Phys. 57  (1977), Issue 3, 882–900}.
\label{bib:Shibata}

\bibitem{Wan-Lei}
Wan-Lei Guo, Cheng-Jun Xia, Tao Lin, and Zhi-Min Wang, 
\textit{Exploring detection of nuclearites in a large liquid scintillator neutrino detector},
\href{https://doi.org/10.1103/PhysRevD.95.015010}{Phys. Rev. D 95 (2017), 015010}.
\label{bib:Wan-Lei}


\bibitem{MACRO}
 M. Ambrosio [MACRO Collaboration],
\textit{Nuclearite search with the MACRO detector at Gran Sasso},
\href{https://doi.org/10.1007/s100520000142}{Eur. Phys. J. C 13 (2000), 453–458 }.
\label{bib:MACRO}

\bibitem{SLIM}
S. Balestra et al,
\textit{Results of the search for strange quark matter and Qballs with the SLIM experiment},
\href{https://doi.org/10.1140/epjc/s10052-008-0747-7}{Eur. Phys. Jour. C 57 (2008), 525}.
\label{bib:SLIM}

%6
% \bibitem{Popa}
% V. Popa [ANTARES Collaboration], 
% \textit{Very large volume neutrino telescopes as magnetic monopole and nuclearite detectors},
% \href{https://doi.org/10.1016/j.nima.2006.05.179}{Nucl. Instrum. Methods Phys. Res. A: Accel. Spectrom. Detect. Assoc. Equip. 567 (2006), Issue 2, 480-482}.
% \label{bib:Popa}

%5

%\bibitem{acceptance}
%M. Anghinolfi, H. Costantini, K. Fratini, D. Piombo, M. Taiuti
%\textit{New measurement of the angular acceptance of the Antares %Optical Module},
%Antares Internal note, 2008

%\href{https://doi.org/10.1016/j.nima.2006.05.179}{Nuclear Instruments and Methods in Physics Research Section A: Accelerators, Spectrometers, Detectors and Associated Equipment, volume 567, Issue 2, 15 November 2006, Pages 480-482}.
%\label{bib:acceptance}


%7

% \bibitem{Nathalie}
%  J. A. Aguilar [ANTARES Collaboration], 
% \textit{Transmission of light in deep sea water at the site of the Antares neutrino telescope}, 
% \href{https://doi.org/10.1016/j.astropartphys.2004.11.006}{Astropart. Phys. 23 (2005), Issue 1, 131-155}.
% \label{bib:Nathalie}


\bibitem{mupage}
G. Carminati\textcolor{\PRCC}{,} M. Bazzotti\textcolor{\PRCC}{,} A. Margiotta \textcolor{\PRCC}{and} M. Spurio, 
\textit{Atmospheric MUons from PArametric formulas: a fast GEnerator for neutrino telescopes (MUPAGE)}, 
\href{https://doi.org/10.1016/j.cpc.2008.07.014}{Comput. Phys. Commun. 179 (2008), Issue 12, 915-923}.
\label{bib:mupage}

\bibitem{mc}
A. Albert [ANTARES Collaboration], 
\textit{Monte Carlo simulations for the ANTARES underwater neutrino telescope}, 
\href{https://doi.org/10.1088/1475-7516/2021/01/064}{\textcolor{\PRCC}{J. Cosmol. Astropart. Phys.} (2021)}.
\label{bib:mc}

\bibitem{DAQ}
\textcolor{\myModifColor}{J. A. Aguilar} [ANTARES Collaboration], 
\textit{The data acquisition system for the ANTARES Neutrino Telescope}, 
\href{https://doi.org/10.1016/j.nima.2006.09.098}{Nucl. Instrum. Meth. A 570 (2007) 107 [astro-ph/0610029]}.
\label{bib:DAQ} 


 
%8
%\bibitem{MDJong}
%M. de Jong, Antares Internal Note.
%\textit{The triggerefficiency program}, ANTARES-SOFT-2009-001.
%\href{https://doi.org/10.1016/j.astropartphys.2009.02.008}{Astroparticle Physics, volume 31, Issue 4, May 2009, Pages 277-283}.
%\label{bib:MDJong}


%9
\bibitem{Hill}
G. C. Hill \textcolor{\PRCC}{and} K. Rawlins,
\textit{Unbiased cut selection for optimal upper limits in neutrino detectors: the model rejection potential technique},
\href{https://doi.org/10.1016/S0927-6505(02)00240-2}{Astropart. Phys. 19 (2003), Issue 3, 393-402}.
\label{Hill}

%10
\bibitem{FeldmanCousins}
Gary J. Feldman and Robert D. Cousins,
\textit{Unified approach to the classical statistical analysis of small signals},
\href{https://doi.org/10.1103/PhysRevD.57.3873}{Phys. Rev. D 57 (1998), 3873 }.
\label{bib:FeldmanCousins}

%11


\bibitem{ref132}
P. Amram [ANATRES Collaboration],
\textit{The ANTARES optical module},
\href{https://doi.org/10.1016/S0168-9002(01)02026-5}{Nucl. Instrum. Methods Phys. Res. A: Accel. Spectrom. Detect. Assoc. Equip. 484 (2002), 1–3, 369-383}.
\label{bib:ref132}

\bibitem{ref133}
\textcolor{\myModifColor}{J. A. Aguilar} [\textcolor{\PRCC}{ANTARES} Collaboration],
\textit{Transmission of light in deep sea water at the site of the ANTARES},
\href{https://doi.org/10.1016/j.astropartphys.2004.11.006}{Astropart. Phys. 23 (2005) 131-155}.
\label{bib:ref133}


\bibitem{Jihad}
A. Albert et al. (ANTARES Collaboration),
\textit{Search for magnetic monopoles with ten years of the ANTARES neutrino telescope},
\href{https://doi.org/10.1016/j.jheap.2022.03.001}{JHEAp 34 (2022) 1-8}.
\label{bib:Jihad}

\bibitem{systematics}
\textcolor{\myModifColor}{J. A. Aguilar} [\textcolor{\PRCC}{ANTARES} Collaboration],
\textit{Zenith distribution and flux of atmospheric muons measured with the 5-line ANTARES detector},
\href{https://doi.org/10.1016/j.astropartphys.2010.07.001}{Astropart. Phys. 34 (2010) 179-184}.
\label{bib:systematics}

\bibitem{conrad}
J. Conrad, O. Botner, A. Hallgren, and C. Pérez de los Heros
\textit{Including systematic uncertainties in confidence interval construction for Poisson statistics},
\href{https://doi.org/10.1103/PhysRevD.67.012002}{Phys. Rev. D 67, 012002 (2003)}.
\label{bib:conrad}

\end{thebibliography}
